\documentclass{aa}  
\usepackage{natbib}
\usepackage[colorlinks=true, linkcolor=blue, urlcolor=blue, citecolor=blue, pdfborder={0 0 0}]{hyperref}
\bibpunct{(}{)}{;}{a}{}{,} 
\usepackage{tipa} 
\usepackage{graphicx}
\usepackage{txfonts}
\usepackage{lipsum}
\usepackage{caption}
\usepackage[utf8]{inputenc}

\usepackage{subcaption}         
\usepackage{lscape}             
\usepackage{placeins}           
\usepackage{orcidlink} 
\usepackage{breqn} 

\begin{document}

\title{The impact of differential rotation on the stochastic excitation \\ of acoustic modes in solar-like pulsators}
\subtitle{}
 \author{G. Biscarrat  \fnmsep \inst{1} \fnmsep \inst{2} \fnmsep \thanks{G. Biscarrat and L. Bessila equally contributed to this work.}
  \and
    L. Bessila \orcidlink{0009-0007-8721-7657} \inst{1} \fnmsep $^{\star}$
  \and
  S. Mathis\inst{1}
  }

   \institute{Université Paris-Saclay, Université Paris Cité, CEA, CNRS, AIM, Gif-sur-Yvette, F-91191, France\\
   \email{leila.bessila@cea.fr}
   \and 
   Université PSL, Paris, 75006, France}

   \date{Received XXXX, Accepted YYYY}

  \abstract
   {Acoustic modes are excited by turbulent convection in the outer convective envelope of solar-like stars. Observational results from asteroseismic studies show that 44 \% of solar-like stars do not present detectable stochastically-excited acoustic modes. This phenomenon appears to be related to their rotation rate and magnetic activity. Indeed, rotation locally modifies the properties of convection, which in turn influences the stochastic excitation of stellar oscillation modes. In a first paper, we showed that uniform rotation tends to diminish the mode amplitudes significantly. However, convective envelopes in solar-type stars are differentially rotating: the rotation rate difference between mid-latitudes and the equator can go up to 60\%, as shown by recent asteroseismic works.}
   {In this paper, we examine the impact of differential rotation on the stochastic excitation of acoustic modes in solar-like stars.}
   {We provide theoretical predictions for the excitation of acoustic modes in a differentially rotating solar-like star. We use the Rotating Mixing-Length Theory approach to model the local influence of differential rotation on convection. We then estimate the resulting impact on power injection by turbulent convection into oscillation modes numerically, using a combination of the MESA stellar structure and evolution code and GYRE stellar pulsation code.}
   {We show that the power injected in acoustic modes differs by up to $30\%$ for stars with the same mean rotation rate $5 \Omega_{\odot}$ but a distinct differential rotation rate. The excitation of axisymmetric acoustic modes is further inhibited in the anti-solar differential rotation regime where the pole of stars rotates faster than their equator when compared to the uniform rotation case. This could hinder mode detection in such configurations. On the contrary, in solar differential rotation regimes where the equator  rotates faster than the poles, acoustic mode excitation is less inhibited when compared to the uniform rotation case. We also study the influence of the azimuthal order, and show that the amplitudes can differ by up to 30\% for modes with the same order $\ell$ and a different azimuthal order $m$, while the trends for sectoral modes (with $\vert m \vert =l$) is the opposite that the one observed for axisymmetric modes.}
   {This study permits a first prediction of the excitation of acoustic modes as a function of the differential rotation in solar-like pulsators. On the one hand, efficient mode excitation may enable the detection of rotational frequency splittings. On the other hand, weaker-than-expected or stronger-than-expected excitation compared to predictions based on uniform rotation, could provide valuable hints into differential rotation. The results are crucial to interpret data from past and ongoing asteroseismic space missions such as \textit{Kepler}/K2 and TESS and to prepare for PLATO.}


    \maketitle
    

\everymath{\displaystyle}

\def\v{{\boldsymbol{v}}}
\def\u{{\boldsymbol{u}}}
\def\uosc{{\boldsymbol{u}_{\rm osc}}}
\def\ut{{\boldsymbol{U_{\rm t}}}}
\def \Bt{{\boldsymbol{B_t}}}
\def \B{{\boldsymbol{B}}}
\def \Bo{{\boldsymbol{B_0}}}
\def \bosc{{\boldsymbol{b_{\rm osc}}}}
\def \Om{{\boldsymbol{\Omega}}}

\def\et{{\boldsymbol{e_{\theta}}}}
\def \er{{\boldsymbol{e_{r}}}}
\def\ep{{\boldsymbol{e_{\varphi}}}}
\def\ex{{\boldsymbol{e_{x}}}}
\def \ey{{\boldsymbol{e_{y}}}}
\def\ez{{\boldsymbol{e_{z}}}}
\def \di{\boldsymbol{\nabla} \cdot}
\def \nab{\boldsymbol{\nabla}}
\def \rot{\boldsymbol{\nabla} \times}
\newcommand{\adv}[2]{\ensuremath \left(\boldsymbol{#1}\cdot \boldsymbol{\nabla}\right) #2}

\newcommand{\Dd}[2]{\ensuremath\frac{\partial #1}{\partial#2}}
\newcommand{\Dt}[1]{\ensuremath\frac{\partial #1}{\partial t}}
\newcommand{\Dtt}[1]{\ensuremath\frac{\partial^2 #1}{\partial t^2}}
\newcommand{\vc}[1]{\ensuremath\boldsymbol{#1}}


\def \rhoo{(\rho_0 + \rho_t)}
\def \Bot{\B}



\section{Introduction}
\par Acoustic modes are a powerful probe of stellar interiors \citep[see the reviews on helioseismology and asteroseismology by][]{aerts_asteroseismology_2010, garcia_asteroseismology_2019, christensen-dalsgaard_solar_2021}. As pressure fluctuations lead to luminosity variations, acoustic oscillation modes are observed at the surface of solar-like stars using photometric data \citep{michel_corot_2008, chaplin_asteroseismic_2010, huber_hot_2019} from space missions such as CoRoT \citep[Convection, Rotation, and Transits,][]{baglin_corot_2006}, \textit{Kepler}/K2 \citep{borucki_kepler_2008, borucki_kepler_2010, howell_k2_2014} and TESS \citep[Transiting Exoplanet Survey Satellite,][]{ricker_transiting_2014}. Amplitudes of solar-like oscillations result from a balance between stochastic excitation, due to turbulent convection \citep[see e.g.][]{samadi_excitation_2001}, and damping \citep[e.g.][]{grigahcene_convection-pulsation_2005} near the surface of the outer convection zone. The driving mechanism has been investigated in many studies \citep{goldreich_solar_1977, balmforth_solar_1992, samadi_excitation_2001, chaplin_model_2005, belkacem_waves_2008}. It has been shown \citep[see e.g.][]{samadi_excitation_2005} that the main contribution driving the oscillations in solar-like pulsators is the Reynolds stresses source term due to turbulent convection. 

\par However, acoustic modes are not detected in a large fraction of stars with a convective envelope and hence where we expect stochastically excited oscillations \citep[e.g.][]{chaplin_predicting_2011}. \cite{mathur_revisiting_2019} studied a sample of 867 solar-like stars observed with the \textit{Kepler} space mission and found that acoustic modes are detected in only 44\% of the stars. This non-detection could result from high magnetic activity, rapid rotation, or other factors such as metallicity or binarity. Including those physical phenomena in the theoretical modelling of stochastic excitation is then paramount. \\
When it comes to rotation, \cite{belkacem_mode_2009} generalised the formalism by \cite{samadi_excitation_2001} taking into account the perturbation of acoustic modes due to uniform rotation. Previously, in \cite{bessila_impact_2024}, we introduced the direct impact of a uniform rotation on turbulent convection, which results in a diminution of the mode excitation rates for rapid rotation. To do so, we used the prescriptions of the Rotating Mixing-Length Theory (R-MLT) to describe the modification of the root-mean-squared (hereafter r.m.s.) velocity and scale of turbulent convection by rotation \citep[e.g.][]{stevenson_turbulent_1979, augustson_model_2019}. This theory relies on the idea that the mode that maximises the heat transport is dominant in the convective flow according to \cite{malkus_heat_1954}. The whole turbulent convective spectrum is then modelled by this single mode. Despite this simplification, R-MLT prescriptions have been compared successfully to 3D non-linear hydrodynamical numerical simulations in a Cartesian geometry in which the rotation axis is aligned with gravity \citep{barker_theory_2014}. Next, \cite{currie_convection_2020} computed the same type of Cartesian numerical simulations with a tilted rotation axis, allowing to study configurations representative of all latitudes in stars. Again, the prescriptions from \cite{stevenson_turbulent_1979} agree well with the results from numerical simulations and hold over several decades in Rossby numbers at most latitudes. However, as pointed out in \cite{currie_convection_2020}, this single-mode theory fails to capture the anisotropy of heat transport near the equator. In a previous work, we thus generalised the existing formalism of the stochastic excitation of stellar oscillation modes to include uniform rotation by taking into account the local modification of convection by rotation using the R-MLT \citep{bessila_impact_2024}. We have shown that the resulting amplitude of acoustic modes can decrease by up to 70\% for a $20~ \Omega_{\odot}$ rotation rate when compared to the non-rotating case, in agreement with the observational tendency highlighted in \cite{mathur_revisiting_2019}.

\par In this article, we extend this formalism to include differential rotation (hereafter DR). Acoustic oscillations at the Sun's surface detected using helioseismology have revealed its rotation profile in the convective zone \citep{chaplin_skew-symmetric_1999, schou_helioseismic_1998, thompson_internal_2003, couvidat_rotation_2003, garcia_tracking_2007}. The latitudinal DR rate between the poles and the equator in the convective envelope is nearly 30\%, whereas the radiative core exhibits a solid-body rotation until $0.25 R_{\odot}$. Other ways exist to access information on the rotation of stars other than the Sun. For instance, spectroscopic measurements can estimate surface velocity and the related differential rotation by observing the broadening of absorption lines caused by the Doppler effect \citep{donati_differential_1997, barnes_dependence_2005}. The photometric variability of starspots at different latitudes can also be used \citep{olah_multiple_2009}. More recently, significant progress has been made in detecting the magnitude of the latitudinal DR by analysing its impact on rotational frequency-splittings of mixed and acoustic modes \citep[see e.g.][]{deheuvels_astrophysics_2015, benomar_nearly_2015, benomar_asteroseismic_2018, bazot_latitudinal_2019}. For example, \cite{benomar_asteroseismic_2018} detected latitudinal DR in some solar-like stars, with a median value of 64 \% difference between mid-latitudes and the equator. These unexpectedly high value for the shear poses a challenge to understanding the angular momentum transport in the convective envelope of low-mass stars \citep[e.g.][]{kapyla_confirmation_2014, brun_powering_2022} and highlight that DR should not be neglected or treated as a small perturbation in theoretical models.

\par Characterising the DR in the convective envelope of solar-type stars is key as it is one of the most important ingredients for dynamo action \citep[e.g.][]{saar_time_1999, charbonneau_dynamo_2005, jouve_exploring_2010, brun_magnetism_2017,noraz_impact_2022}. For a given star, the DR profile is directly related to the dimensionless fluid Rossby number $\mathcal{R}o_{\rm f}$ (defined in Eq. \ref{eq:def_rof}), which quantifies the ratio between the inertia term and the Coriolis acceleration in Navier-Stokes equation  \citep[see e.g.][]{gilman_nonlinear_1977, gilman_compressible_1981, matt_convection_2011, guerrero_differential_2013, gastine_solar-like_2014, kapyla_confirmation_2014, brun_differential_2017, brun_powering_2022, noraz_impact_2022}. Studies using direct numerical simulations highlight a change in the DR profile between intermediate and slow rotators. The transition between these two states occurs around $\mathcal{R}o_{\rm f} = 1$ in numerical simulations \citep{gastine_solar-like_2014, kapyla_confirmation_2014, karak_magnetically_2015,  brun_powering_2022}. Intermediate rotators ($0.15 \lesssim \mathcal{R}o_{\rm f} \lesssim 1$) exhibit a solar-like conical DR characterized by a slow rotation at the poles and a faster rotation at the equator \citep[e.g.][]{brun_differential_2017}. In contrast, slow rotators ($\mathcal{R}o_{\rm f} \gtrsim 1$) are the seat of an anti-solar DR, with a slower equator and faster poles. Following these numerical results, \cite{benomar_asteroseismic_2018} analysed the DR in 40 solar-like stars using asteroseismology and found that some stars could have an anti-solar DR profile. Doppler imaging spectroscopy allowed some other detections in subgiants and red giants stars \citep[see e.g.][]{kovari_surface_2007}. \cite{noraz_hunting_2022} also identified main-sequence stars potentially hosting anti-solar DR in the \textit{Kepler} sample. 
Finally, several trends for the DR in solar-like stars have been uncovered by numerical simulations and observations: the DR $\Delta \Omega$ appears to increase with the stellar mass. Numerical simulations also revealed dependencies between the DR and the rotation: $\Delta \Omega \sim \Omega_{\star}^r$, with $r$ a positive exponent between $0.2$ and $0.7$ \citep[e.g.][]{reinhold_fast_2013}. However, the exact value of the exponent $r$ is still under investigation \citep[e.g.][]{brun_powering_2022}.

\par In this framework, since we have demonstrated how a uniform rotation can impact the injection of energy by turbulent convection into acoustic modes propagating in solar-type stars \citep{bessila_impact_2024}, their potentially strong DR must be taken into account. In this work, we aim to understand how acoustic mode amplitudes are affected by the DR of a star through the modification it induces on turbulent convection that in turn excites these modes. Since acoustic modes are the ones enabling to characterise the DR in the convective envelope of low-mass stars, it is indeed crucial to coherently predict their excitation rate and amplitude. This will then allow us to better constrain the observed DR, which is key to understanding magnetism and angular momentum redistribution in these regions. This paper is organised as follows: we present the theoretical model of acoustic mode excitation in a differentially rotating convective envelope in Section \ref{sec:model}. We then implement the obtained theoretical prescriptions to estimate the energy injected into acoustic oscillation modes in a differentially rotating solar-type star modelled using the MESA and GYRE stellar modelling suite in Section \ref{sec:numerical_estimates}. Section \ref{sec:star_mass} explores the influence of stellar mass on DR and its subsequent effect on the excitation of acoustic modes. Finally, we present the key results of our study and discuss their implications for understanding observed acoustic mode amplitudes in differentially rotating solar-like stars in Section \ref{sec:conclusion}.


\section{Turbulent stochastic excitation in differentially rotating stars}
\label{sec:model}

\subsection{Differential rotation}

Helioseismology has demonstrated that the Sun's convective zone rotates nearly uniformly in the radial direction \citep[see e.g.][]{schou_helioseismic_1998, thompson_internal_2003, garcia_tracking_2007, eff-darwich_dynamics_2013}. In addition, stars exhibit different DR profiles depending on their evolutionary stage \citep{noraz_magnetochronology_2024}. First, the DR profile is mostly cylindrical during the Pre-Main Sequence (when $\mathcal{R}o_{\rm f} \lesssim 0.15$). During this phase, low-mass stars are fast rotators, and the Coriolis acceleration constrains the flows because of the Taylor-Proudman constraint \citep{proudman_motion_1916, taylor_motion_1917}. In this case, the local rotation rate depends on the distance to the rotation axis $\Omega(r,\theta) = \Omega(r \sin \theta)$, where $\theta$ is the colatitude and $r$ the radius \citep[see e.g.][]{brun_differential_2017, brun_powering_2022}. During the Main Sequence, the DR becomes conical (when $\mathcal{R}o_{\rm f} \gtrsim 0.15$): the local rotation rate depends mainly on the colatitude, as in the Sun, where the equator rotates faster than the poles. For more massive or evolved stars, the conical DR can become antisolar if $\mathcal{R}o_{\rm f} \gtrsim 1$ with stellar poles rotating faster than the equator. As \textit{Kepler} observational sample is dominated by Main-Sequence stars, we thus focus in this study on a conical DR profile, to provide a tractable model, that can easily be applied to a broad sample of stars. We model the rotation profile as follows:

\begin{equation}
    \Omega (\theta) = \Omega_{\rm p} + \Delta \Omega \sin^2\theta,
    \label{eq:conic_diff}
\end{equation}

\noindent where $\Omega_{\rm p}$ is the rotation rate at the pole and $\Delta \Omega \equiv \Omega_{\rm eq} - \Omega_{\rm p}$ is the difference between the rotation rate at the equator and the poles. If $\Delta \Omega > 0$, the rotation rate at the equator is higher than the one at the poles: it is a solar DR profile. On the contrary, if $\Delta \Omega < 0$, the rotation profile is antisolar. As we wish to compare the impact of DR to the uniformly rotating case, we define the mean rotation rate over the convective zone as:

\begin{equation}
    \tilde{\Omega} = \int_0^{\pi/2} \Omega(\theta) \sin \theta d \theta = \frac{2 \Omega_{\rm eq} + \Omega_{\rm p}}{3}.
\label{eq:mean_omega}
\end{equation}
We then chose $\tilde{\Omega}$ as the reference rotation rate in our model.

\begin{figure}[!h]
    \centering
\includegraphics[width=0.5\textwidth]{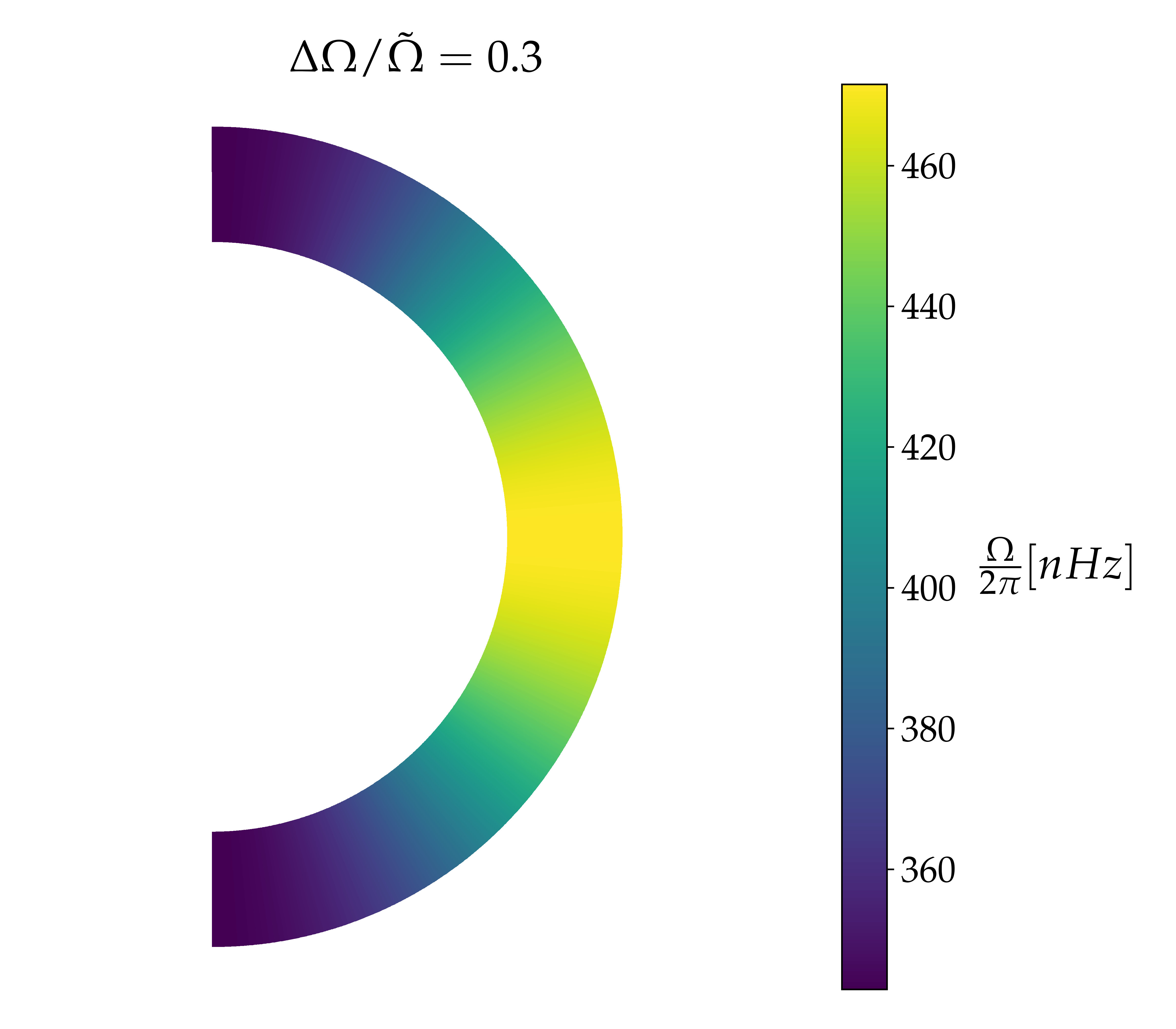}
\includegraphics[width=0.5\textwidth]{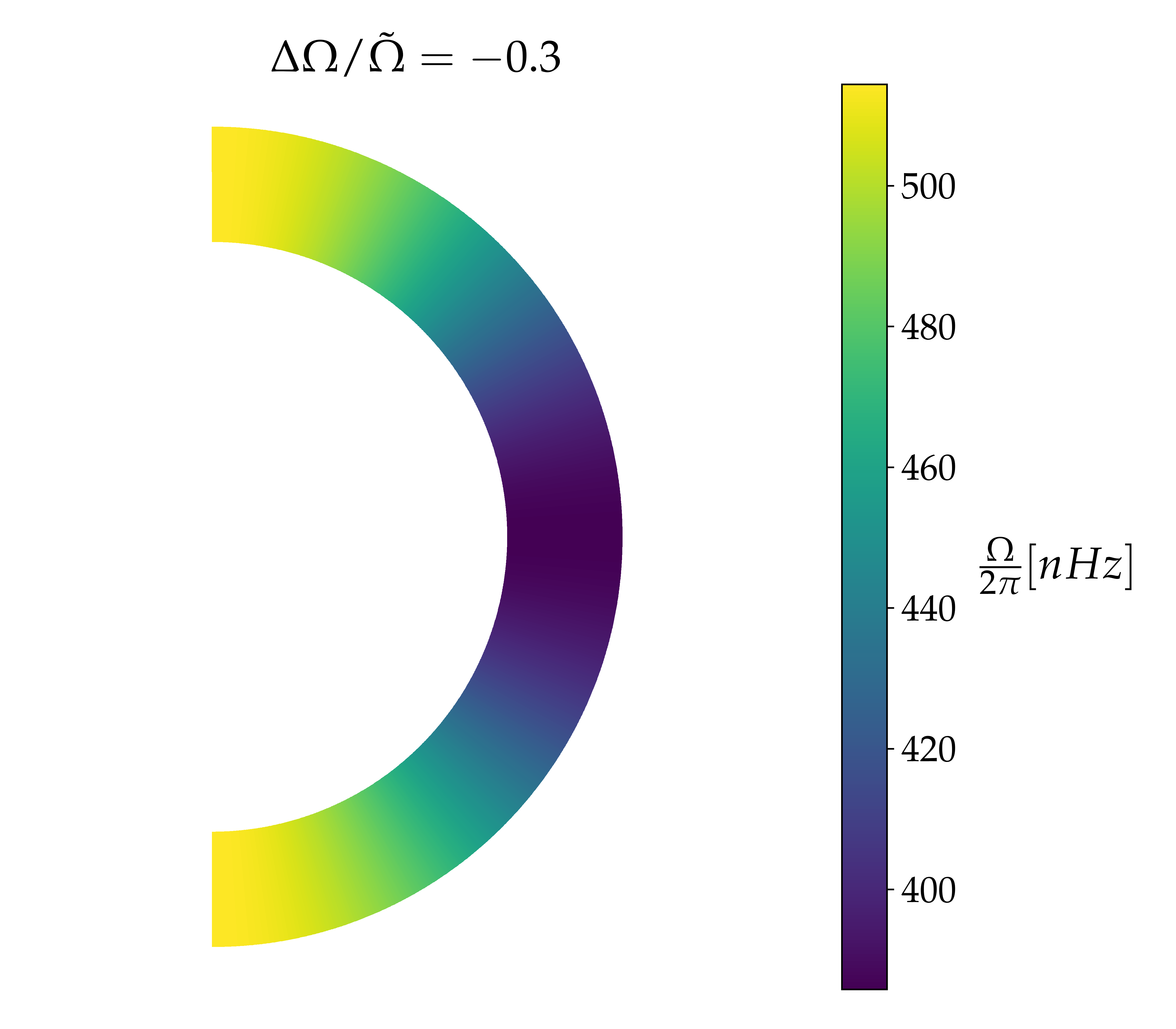}
    \caption{Conical DR profile for a solar regime (Top), and an anti-solar regime (Bottom), following Eq. (\ref{eq:conic_diff}). In both cases $\tilde{\Omega}/2 \pi = 417 ~ \rm nHz$.}
    \label{fig:conical_dr}
\end{figure}

According to the observational study by \cite{benomar_asteroseismic_2018}, the relative DR rate between the equator and mid-latitude can go up to $(\Omega_{45}- \Omega_{\rm eq})/\Omega_{\rm eq}\sim 0.6$, which correspond to $\Delta \Omega/\tilde{\Omega} \sim 2$ in our model, following Eq. (\ref{eq:conic_diff}) and (\ref{eq:mean_omega}). 
For simplicity, we define the relative DR rate: 
\begin{equation}
    A = \frac{\Delta \Omega}{\tilde{\Omega}},
\end{equation}
so that the DR profile writes: 
\begin{equation}
    \Omega(\theta) = \tilde{\Omega} \left(1 + A \left(\sin^2 \theta - \frac{2}{3} \right)\right).
\end{equation}
We plot in Fig. \ref{fig:conical_dr} the conical DR profile, where we chose the solar value for the mean rotation rate: $\tilde{\Omega}_{\odot}/2 \pi = (2 \Omega_{\rm eq,\odot} + \Omega_{\rm p,\odot})/6 \pi = 417~ \rm nHz$, where $\Omega_{\rm eq, \odot}$ and $\Omega_{\rm p, \odot}$ are the solar rotation rates at the equator and the poles, respectively.

\subsection{Excitation by turbulent convection}
We build on the formalisms by \cite{samadi_excitation_2001}, \cite{belkacem_mode_2009} and \cite{bessila_impact_2024}. We only take into account the Reynolds stresses in the convective envelope as the source that drives the oscillation modes in the stellar cavity. Indeed, \cite{samadi_excitation_2007} showed that the entropy fluctuations account for only 10 \% of the total injected power, and \cite{belkacem_mode_2009} demonstrated that the source terms coming from the Coriolis acceleration and the DR are negligible as well. Nonetheless, DR plays a role in locally modifying convective properties, thereby influencing the Reynolds stresses as a source term. Linearising Navier-Stokes and continuity equations, then separating the velocity due to the oscillations $\uosc$ from the one due to turbulent convection $\ut$, we derive the forced wave equation \citep[we refer the reader to][for more details about this derivation]{belkacem_mode_2009}:

\begin{equation}
\rho_0 \left(\frac{\partial^2}{\partial t^2} - \mathcal{L}\right)\uosc + \mathcal{D} = \Dt{\mathcal{S}_R} + \Dt{\mathcal{C}},
    \label{eq:inhomogeneous}
\end{equation}

\noindent where $\rho_0$ is the density, $\mathcal{L}$ is the linear operator describing the propagation of waves, $\mathcal{D}$ is the damping term and $\partial \mathcal{C}/\partial t$ contains negligible source terms. $\mathcal{S}_R$ is the Reynolds-stresses source term defined as:
\begin{equation}
    \Dt{\mathcal{S}_R} = -\frac{\partial}{\partial t} \Bigg[ \nabla: \Big( \rho_0 \ut \ut \Big)\Bigg],
\end{equation}
which acts as the forcing for the waves.

\par The wave velocity field is related to the displacement by the expression:
\begin{equation}
\uosc = A(t) \left[ i\widehat{\omega} \boldsymbol{\xi}(\vc{r}) - (\boldsymbol{\xi}(\vc{r}) \cdot \nabla \boldsymbol{\Omega}) r \sin \theta \, \boldsymbol{e}_{\phi} \right] e^{i \omega_0 t},
\label{eq:velocity_osc}
\end{equation}

\noindent where $\widehat{\omega} = \omega_0 + m\Omega$, and $\omega_0$ is the mode frequency in the inertial frame without rotation, $\vc{\xi}(\vc{r})$ is the adiabatic Lagrangian displacement of the waves (i.e. with no forcing), and $A(t)$ is the instantaneous amplitude due to the turbulent forcing \citep{unno_non-radial_1989}. In this work, we consider purely radial modes, with a conical DR profile that only depends on the colatitude $\theta$. In this framework, the second term in the right-hand side of Eq. (\ref{eq:velocity_osc}) does not contribute. Moreover, we consider acoustic modes with high frequencies, which are weakly perturbed by the rotation such that $\omega_0 \gg \vert m \vert \Omega$.  The wave velocity becomes: 

\begin{equation}
\uosc =  i \omega_0 A(t) \boldsymbol{\xi} \, e^{i \omega_0 t},
\label{eq:velocity}
\end{equation}

\noindent In this framework, for a mode of given $\ell, m$, the displacement function $\vc{\xi}(\vc{r})$ is expanded on  the vectorial spherical harmonics \citep[e.g.][]{unno_non-radial_1989}: 

\begin{equation}
    \vec{\xi}(\vec{r})= \left[\xi_{r ; n, \ell, m} (r) \vec{e}_r 
    + \xi_{h ; n, \ell, m} (r)\vec{\nabla}_h 
    + \xi_{t ; n, \ell, m}(r) \vec{\nabla}_h \times \vec{e}_r \right] 
    Y_{\ell, m} (\theta, \varphi),
\end{equation}

\noindent where $\xi_r, \xi_h$, and $\xi_t$ are respectively the radial, horizontal, and toroidal components of the displacement eigenfunctions. We introduced the spherical harmonics $Y_{\ell, m} (\theta, \varphi)$.  We have also introduced the horizontal gradient: 
\begin{equation}
    \vec{\nabla}_h = \frac{\partial}{\partial \theta} \vc{e}_{\theta} + \frac{1}{\sin \theta} \frac{\partial}{\partial \phi} \vc{e}_{\phi}.
\end{equation}

\par Using the forced wave equation (\ref{eq:inhomogeneous}) and Eq. (\ref{eq:velocity}), we find the mean square amplitude $\langle \lvert A(t) \rvert^2 \rangle$ of $\uosc$, which is directly related to the power $\mathcal{P}$ injected into each mode, of given radial, latitudinal, and azimuthal order $(n,\ell,m)$ \citep{samadi_excitation_2001, belkacem_mode_2009} through:

\begin{equation}
    \mathcal{P} = \eta \langle \lvert A \rvert ^2 \rangle I \omega_0^2, 
\end{equation}
\noindent where $\eta$ is the damping rate of the studied mode. We introduce the mode inertia: 
\begin{equation}
    I = \iiint_{\mathcal{V}} \rho_{0} d^{3} r\left(\boldsymbol{\xi}^{*} \cdot \boldsymbol{\xi}\right) = \int_{0}^{R_\star} \left[ \xi_r^2 + \ell (\ell + 1) \xi_h^2 \right]\rho_{0} d r,
\end{equation}
where the exponent $^*$ denotes the complex conjugate, $R_\star$ is the stellar radius, and we perform the integral over the whole star volume $\mathcal{V}$. As we study high-frequency acoustic modes, we can neglect the inertia correction introduced in the rapid rotation regime by \cite{neiner_astronomy_2020}.
In addition, we focus in this work on modes with frequencies between 1 mHz and 5 mHz and low-angular degree $\ell$ modes, where $\ell \ll n$. Such modes are essentially radial, i.e. $\xi_r \gg \xi_h$ \citep[e.g.][]{belkacem_stochastic_2008}. We then assume purely radial modes as a first step. 
Following the derivation in \cite{samadi_excitation_2001}, \cite{belkacem_mode_2009} and \cite{bessila_impact_2024} under this assumption, we find:

\begin{equation}
\left\langle|A|^2\right\rangle=\frac{1}{8 \eta\left(\omega_0 I\right)^2}C_R, 
\end{equation}
where $C_R$ is the Reynolds-stress contribution: 

\begin{equation}
    C_R=\frac{16}{15} \pi^3 \int_{\mathcal{V}} d^3 x_0 \rho_0^2\left|\frac{\mathrm{d} \xi_r} {\mathrm{~d} r}\right|^2 Y_{\ell,m} Y^{*}_{\ell,m}\hat{S}_R\left(r,\theta,\omega_0\right), 
    \label{eq:reynolds-int}
\end{equation}
where we define the Reynolds stresses source contribution:

\begin{equation}
\hat{S}_R\left(\omega_0\right)=\int \frac{d k}{k^2} E^2(k) \int d \omega \chi_k\left(\omega+\omega_0\right) \chi_k(\omega). 
\label{eq:s_hat_def}
\end{equation}

\noindent We introduce $E(k)$ the spatial kinetic energy spectrum of turbulence, and $\chi_k(\omega)$ the associated eddy time-correlation function following \cite{stein_generation_1967}. Although rotating turbulence is anisotropic \citep[see e.g.][]{ecke_turbulent_2023}, we make the simplification of isotropic turbulence. This approximation is commonly adopted to have a more tractable model while effectively capturing the dynamics of the stochastic excitation occurring at small scales \citep[e.g.][]{samadi_excitation_2001}. As demonstrated in our previous work \citep{bessila_impact_2024}, rotation does not significantly impact mode excitation when a Gaussian eddy time-correlation function is assumed. However, observations indicate that mode amplitudes depend on rotation \citep[e.g.][]{mathur_revisiting_2019}. When choosing a Lorentzain eddy-time correlation, the injected power becomes sensitive to rotation \citep{bessila_impact_2024}, suggesting that a Lorentzian spectrum provides a more accurate representation. Additionally, a Lorentzian time-correlation spectrum shows better agreement with solar observations \citep[e.g.][]{belkacem_turbulent_2010}. We therefore model the eddy time-correlation spectrum using a Lorentzian function:
\begin{equation}
\displaystyle
\chi_k(\omega)=\frac{1}{\pi \omega_k} \frac{1}{1+\left(\frac{\omega}{\omega_k}\right)^2},
\end{equation}
where $\omega_k$ is the frequency of an eddy of wavenumber $k$.
When it comes to the spatial kinetic energy spectrum $E(k)$, we chose a Kolmogorov spectrum: $E(k) \propto k^{-5/3}$ \citep{kolmogorov_dissipation_1941}. More choices for the turbulent spectra, $E(k) \propto k^{-\alpha}$, in rotating convection have been discussed in our previous study \citep{bessila_impact_2024}. We highlighted that higher values for $\alpha$ lead to higher values for the resulting power injected into the acoustic modes, although the global tendency for the power injection does not change depending on this choice.

\par
\begin{figure}[]
    \centering
\includegraphics[width=0.5\textwidth]{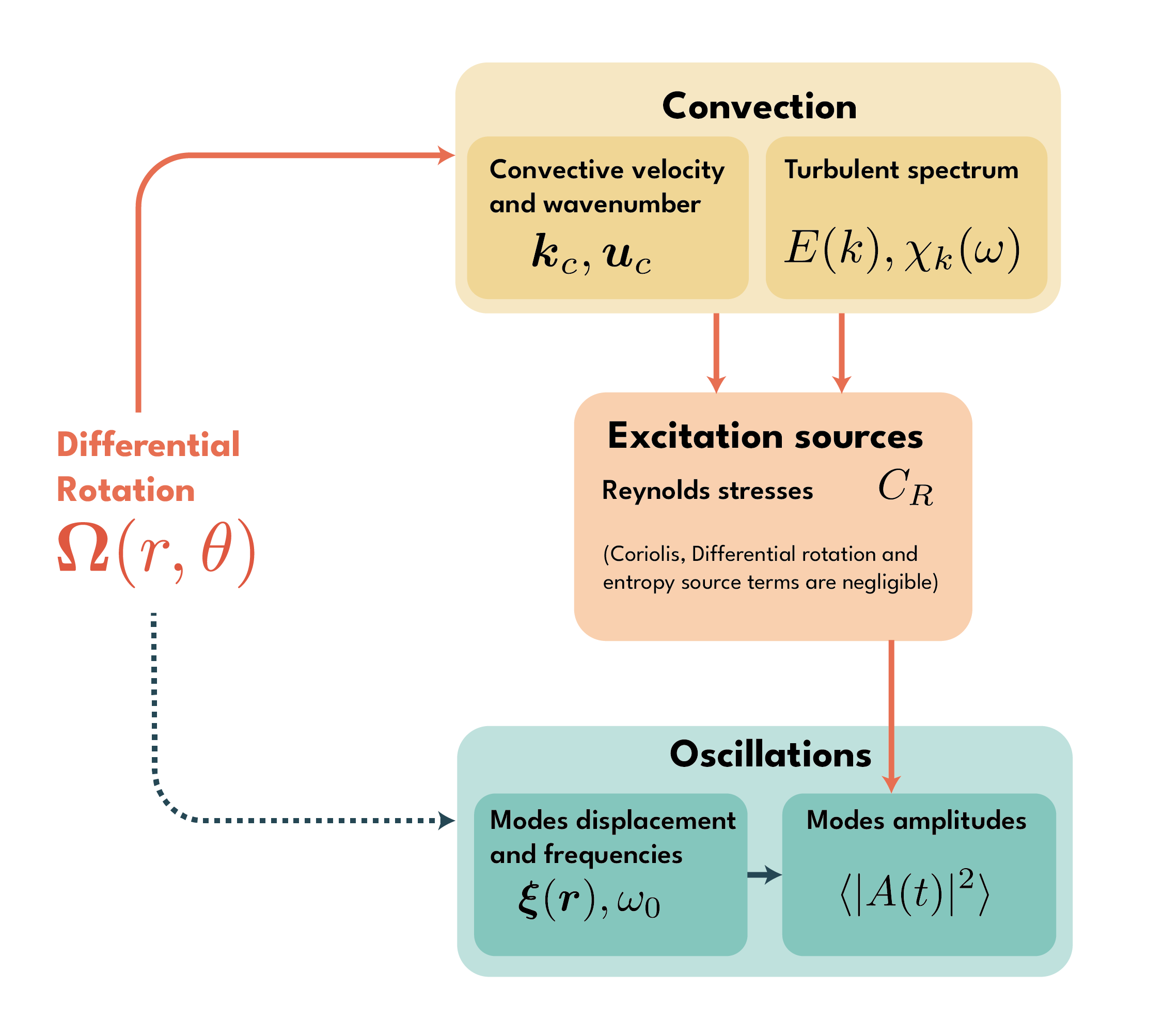}
    \caption{Interdependencies in the stochastic excitation formalism with DR. The red arrows denote the direct impact of differential rotation on convection we take into account, while the dotted ones represent phenomena we do not take into account. Dark arrows represent the general interdependencies in the stochastic excitation formalism.}
    \label{fig:interdependencies}
\end{figure}

 In this framework, the power injected into a given acoustic mode is:

\begin{equation}
    \mathcal{P} = \frac{C_R}{8 I^2}.
    \label{eq:power_formula}
\end{equation}
\noindent Even if uniform and differential rotation do not add directly any non-negligible source term in this formalism, convection is modified by the local rotation and thus by the DR \citep[e.g.][]{stevenson_turbulent_1979, augustson_model_2019}, hence indirectly influencing mode excitation. Fig. \ref{fig:interdependencies} sums up the complex interplay between waves, convection, DR and stochastic mode excitation. 

\subsection{Rotating convection}
\label{sub:rotating_convection}

\par To model the local influence of differential rotation on turbulent convection, we make use of the prescription from \cite{augustson_model_2019} of the Rotating Mixing-Length Theory (hereafter R-MLT). This formalism is a modified version of the Mixing-Length Theory \citep[e.g.][]{bohm-vitense_uber_1958}, taking into account the effects of the Coriolis acceleration. It is a single-mode approach that follows the heat flux maximisation principle proposed by \cite{malkus_heat_1954}: the convective flow is dominated and modelled by the mode that transports the most heat. It has been successfully compared with local direct numerical simulations in a Cartesian geometry \citep[see e.g.][]{barker_theory_2014, currie_convection_2020}, as well as with global Large-Eddy Simulations for the prediction of the convective penetration \citep{korre_dynamics_2021}. It also has given some insightful results for the modelling of light elements mixing in low-mass stars \citep{dumont_lithium_2021} and the thermal properties of the convective core boundary in early-type stars \citep{michielsen_probing_2019}.
\cite{augustson_model_2019} do not account for the term $r \sin \theta (\vc{u} \cdot \nabla \vc{\Omega}) \hat{\vc{e}}_\varphi$ in the momentum equation in their model \citep[e.g.][]{unno_non-radial_1989}. We thus consider convective length scales which are much smaller than those associated with the latitudinal variation of the angular velocity. By doing so, we treat locally the impact of differential rotation on convection.

\par The convective velocity (resp. convective wavenumber) with rotation, $u_c$ (resp. $k_c$), is expressed as a function of the convective velocity (resp. convective wavenumber) without rotation, $u_0$ (resp. $k_0$): 
 \begin{equation}
\begin{aligned}
u_c = \tilde{U}(\mathcal{R}o) u_0, \\
k_c = \tilde{K}(\mathcal{R}o) k_0,
\end{aligned}
\end{equation}
where we introduce the Rossby number following \cite{stevenson_turbulent_1979}, which represents the ratio between the advection and Coriolis acceleration: 
\begin{equation}
    \mathcal{R}o = \frac{u_0 k_0}{2 \Omega \cos{\theta}}, 
    \label{eq:rossby_def}
\end{equation}
$\theta$ being the colatitude. In this definition, the Rossby number represents the ratio between the convective turnover frequency and the frequency associated with the vertical component of the rotation vector. \cite{augustson_model_2019} have shown that only the local vertical component of the rotation vector influences convection when assuming the principle of heat-flux maximization in which the  heat is transported by the most unstable mode within the rotating mixing-length theory. Note that in other studies, such as \citet{augustson_model_2019}, the Rossby number is defined without the explicit $\cos \theta$ dependence. In this cases, the latitudinal variation of the rotation vector is instead incorporated directly into the governing equations.

In the formalism by \cite{augustson_model_2019}, one can find the velocity modulation and the wavenumber modulation: 
\begin{equation}
    \tilde{U} = \frac{5 \tilde{s}}{\sqrt{6}} z^{-1/2},
\label{eq:u_tilde}
\end{equation}

\begin{equation}
    \tilde{K} = \sqrt{\frac{2}{5}} z^{3/2},
\label{eq:k_tilde}
\end{equation}

\noindent where $z$ is the solution of the polynomial equation corresponding to Eq. (46) in \cite{augustson_model_2019}:

\begin{equation}
    2 z^5-5 z^2-\frac{18 }{25 \pi^2 \mathcal{R}o^2 \tilde{s}^2}=0,
\end{equation}
where $\tilde{s}$ is a dimensionless prefactor defined as $ \tilde{s} = 2^{1/3} 3^{1/2} 5^{-5/6} $. We refer the reader to Appendix A of \cite{bessila_impact_2024}, where we detailed the R-MLT model from \cite{augustson_model_2019}.

\section{Impact of the differential rotation on the excitation of acoustic modes}
\label{sec:numerical_estimates}

\begin{figure*}[]
    \centering
    \includegraphics[width=0.7\textwidth]{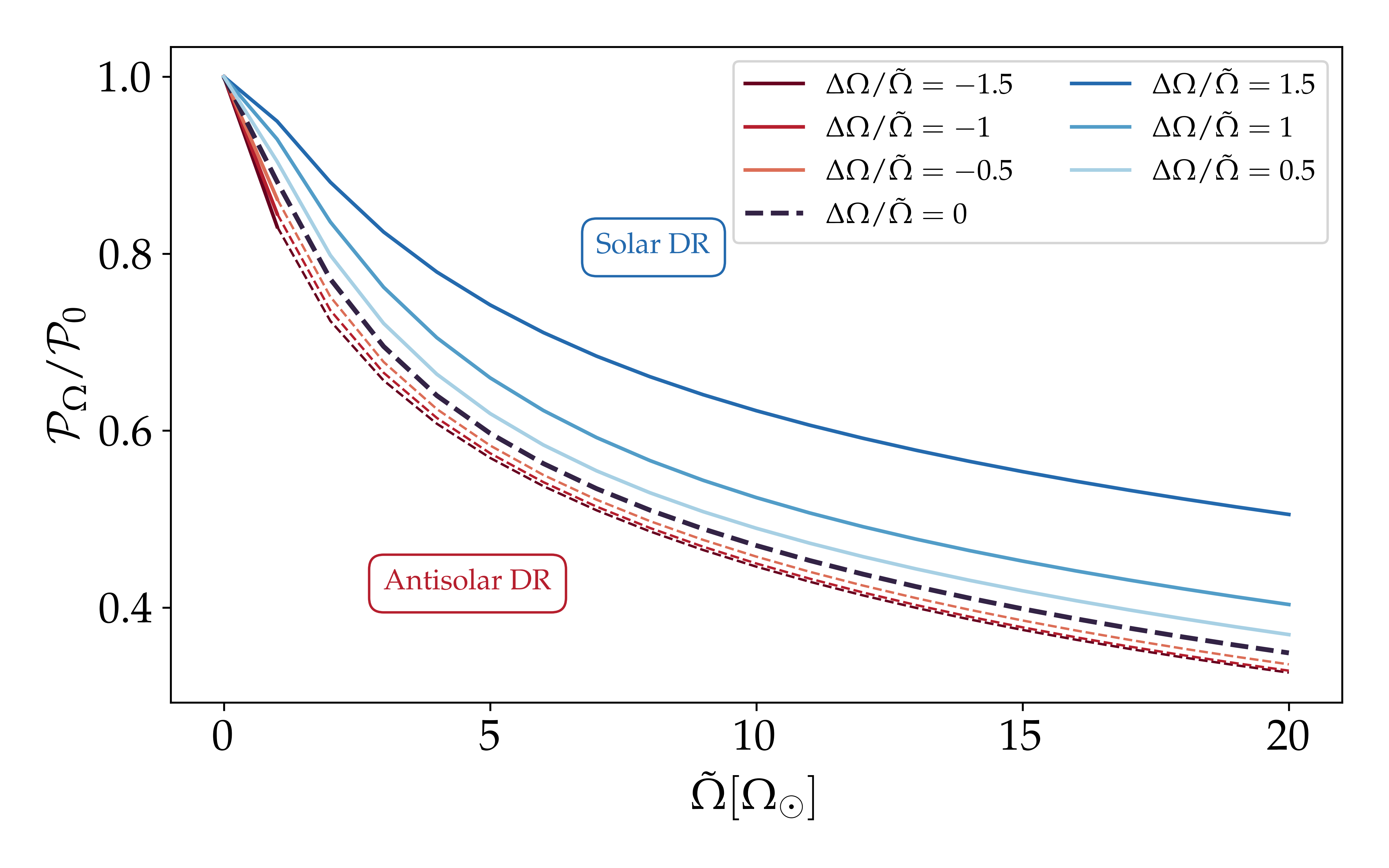}
    \caption{Influence of the DR on the power injected in the acoustic mode ($n=7, \ell = 0, m=0$) in a $1 M_{\odot}, Z = 0.02$ Solar-like model. Anti-solar regimes, characterised by $\Delta \Omega / \tilde{\Omega} < 0$, are represented in red, whereas solar DR regimes are drawn in blue. The uniform rotation regime is drawn with a thicker dashed black line. For $\tilde{\Omega} > 1 \Omega_{\odot}$, such a Solar-like star is unlikely to present an anti-solar DR, because $\mathcal{R}o_{\rm f} < 1$. We thus represented this regime with dashed lines.}
    \label{fig:differential_rotation}
\end{figure*}

\begin{figure*}[]
    \centering    
    \includegraphics[width=0.99\textwidth]{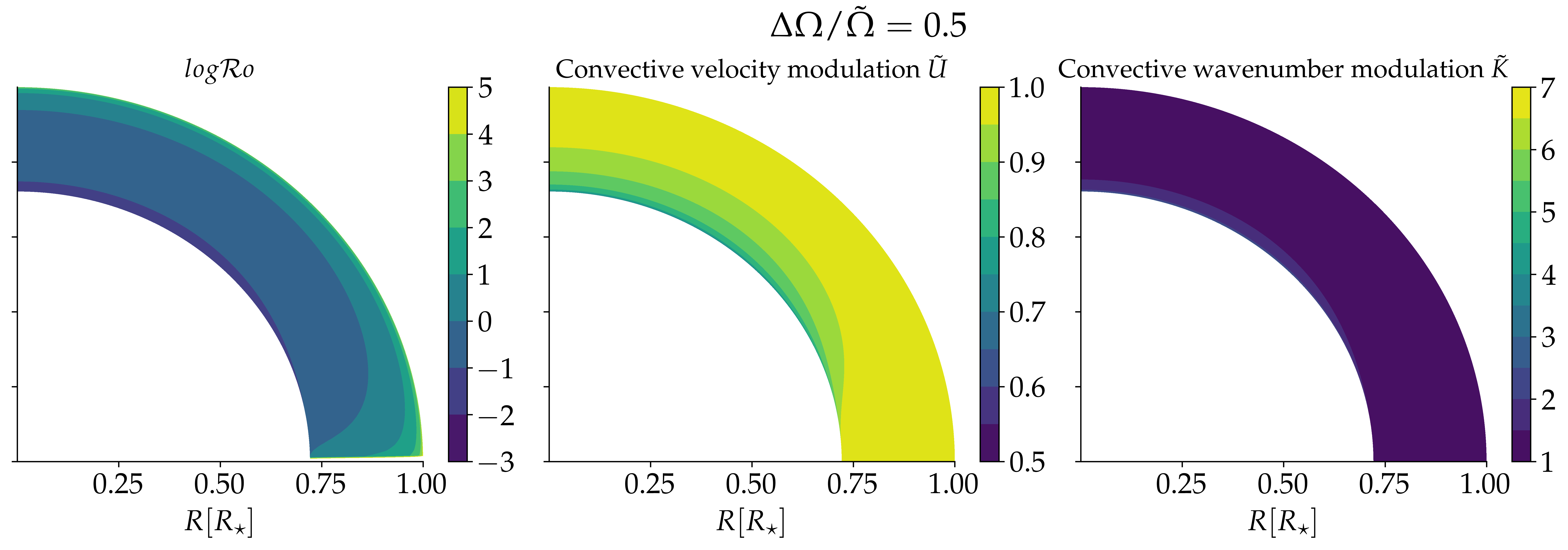}
    
    \includegraphics[width=0.99\textwidth]{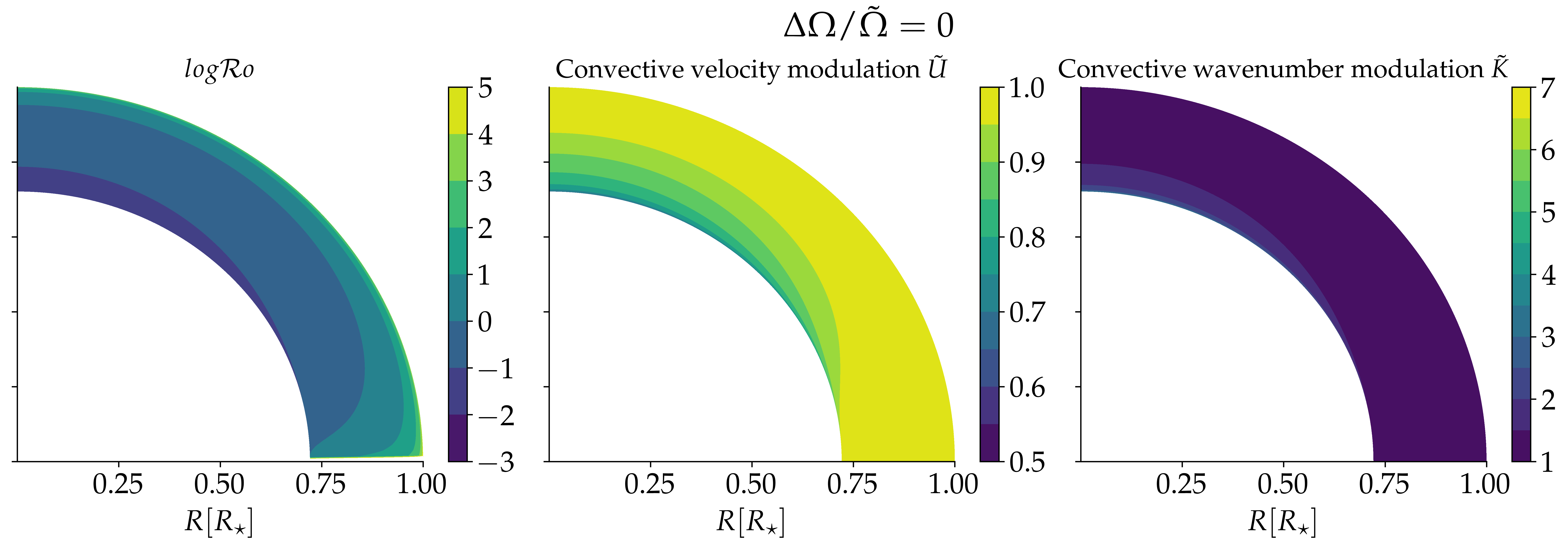}
    
    \includegraphics[width=0.99\textwidth]{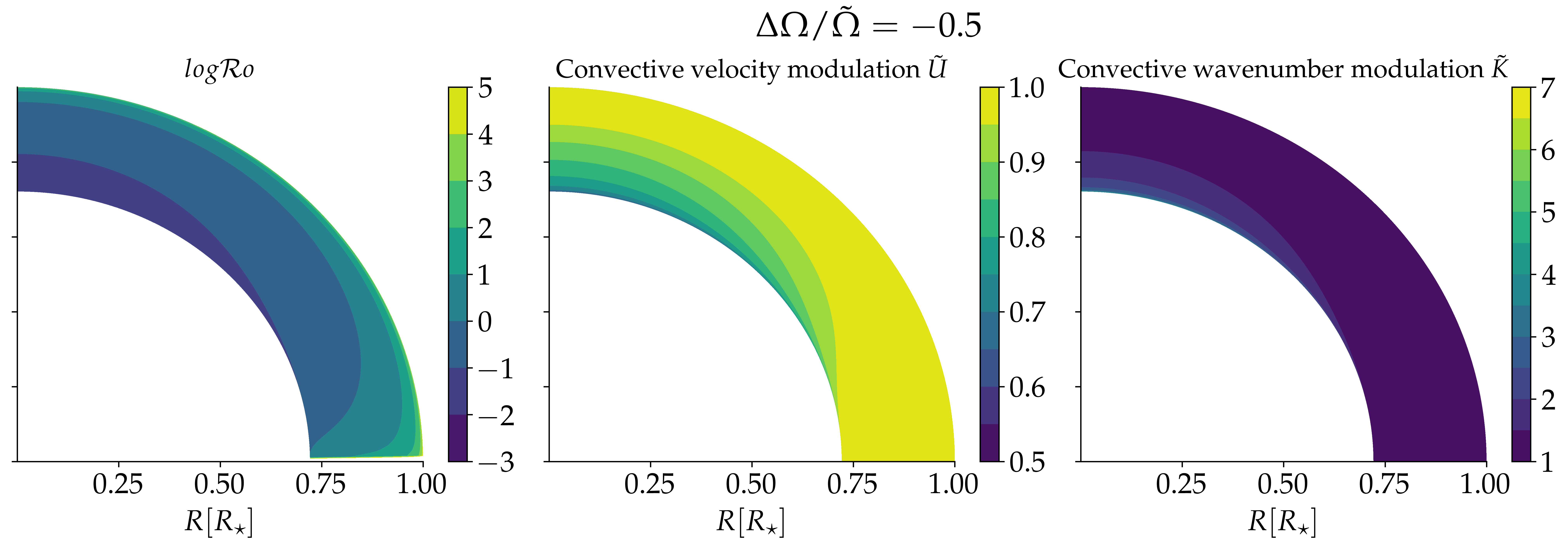}
    
    \caption{Modification of the Rossby number (left panel), convective velocity (centre panel) and convective wavenumber (right panel) following \cite{augustson_model_2019}, for a fixed mean rotation rate $\tilde{\Omega} = 1 \tilde{\Omega}_{\odot}$ corresponding to a 27 days period, and different values of latitudinal DR: $\Delta \Omega /\tilde{\Omega} = +0.5$ (top panel), $\Delta \Omega /\tilde{\Omega} = 0$ (middle panel), $\Delta \Omega /\tilde{\Omega} = -0.5$ (bottom panel).}
    \label{fig:spheres_soleil}
\end{figure*}

\subsection{Method}

In this section, we compute the power injected by turbulent convection into the acoustic modes as a function of the DR. We follow the theoretical derivation expounded upon above. We use a Solar-like $1 M_{\odot}, Z = 0.02$ 1D model, computed with the stellar evolution code MESA \citep{paxton_modules_2011, paxton_modules_2013, paxton_modules_2015, paxton_modules_2018, paxton_modules_2019, jermyn_modules_2023} to evaluate convective non-rotating velocity and scale ($u_0,k_0)$ and thermodynamical quantities (i.e. pressure and density). We use the stellar oscillation code GYRE \citep{townsend_gyre_2013, townsend_gyre_2014} to compute stellar acoustic radial modes eigenfunctions $\boldsymbol{\xi}_{r; n, \ell, m}$ and their eigenfrequencies. 
For each mode, we use the following methodology: 
\begin{enumerate}
    \item We define the DR rate $\Delta \Omega$ and compute the resulting rotation profile.
    \item For a given value of the local rotation $\Omega$ at a given radius and latitude, we compute the local effective Rossby number using the convective non-rotating properties $u_0$ and $k_0$ from MESA. 
    \item We then compute the convective velocity and convective wavenumber modified by rotation, using \cite{augustson_model_2019} prescriptions for the R-MLT (Equations \ref{eq:u_tilde} and \ref{eq:k_tilde}).
    \item Finally, we compute the resulting power injected into the modes, following the set of equations (\ref{eq:power_formula}), (\ref{eq:reynolds-int}) and (\ref{eq:s_hat_def}).
\end{enumerate}

\noindent In the following work, we take the mean rate of the Sun as a reference: $2 \pi/\tilde{\Omega} = 27$ 
days \citep{thompson_class_2012}.

\subsection{Results}

\begin{figure}
    \centering
    \includegraphics[width=\linewidth]{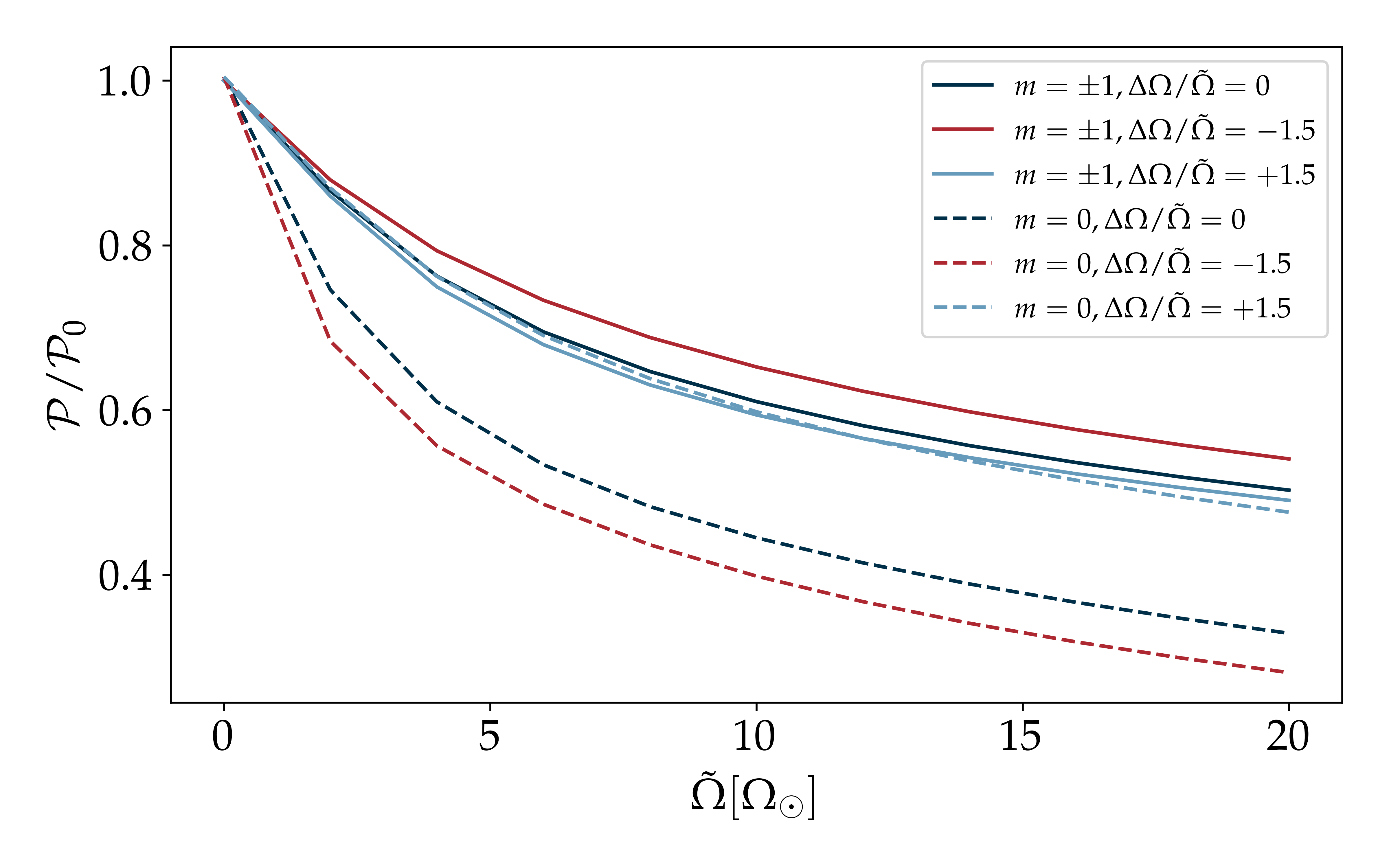}
    \caption{Influence of the azimuthal order $m$ on the power injected by the stochastic excitation for the $\ell = 1$, $n = 7$ mode, for different values of the DR.}
    \label{fig:influence_m_dr}
\end{figure}

\par First, we examine the impact of DR on a fixed acoustic mode $\ell = 0$, $m=0$, $n=7$. In our study of the uniform rotation \citep{bessila_impact_2024}, we demonstrated that this mode's amplitude is among the most significantly inhibited in rapidly rotating stars. Therefore, we focus on this mode in the present work. We use a $1 M_{\odot}$ Solar-like model, with metallicity $Z=0.02$ (see Appendix \ref{sec:inlist} for the MESA inlists). Here, we consider a fixed value for the turbulent spectrum slope: $\alpha = -5/3$ \citep{kolmogorov_dissipation_1941}. For each value of the DR, we compute the power $\mathcal{P}_0$ injected into the oscillations without rotation ($\vc{\Omega}=0$), and the power $\mathcal{P}_{\Omega}$ injected taking DR into account. 

\par Fig. \ref{fig:differential_rotation} shows the ratio $\mathcal{P}_{\Omega}/\mathcal{P}_0$ as a function of rotation for different values of the DR. Rotation tends to reduce the convection strength for the convective mode that transports the most heat in the framework of R-MLT \citep{stevenson_turbulent_1979, augustson_model_2019}: the higher the rotation rate, the weaker the convective velocity with this model, and the lower the power injected by the stochastic excitation. The impact of the DR on the Rossby number, convective velocity modulation $\tilde{U}$ and convective wavenumber modulation $\tilde{K}$ are shown in Fig. \ref{fig:spheres_soleil}.

\par First, we note that for solar DR regimes (i.e. for $\Delta \Omega >0$, in blue in Fig. \ref{fig:differential_rotation}), mode amplitudes are less influenced by the rotation, hence decreasing less rapidly than in the case of uniform rotation when the rotation rate increases. The Rossby number as defined in Eq. (\ref{eq:rossby_def}) scales like $\mathcal{R}o \sim 1/\left(\Omega(\theta)\cos \theta \right)$ and increases towards the equator. It means that rotation has less impact on convection near the equator, than near the poles. For this reason, a rapidly rotating pole has more impact on convection than a slowly rotating pole. As a consequence, in the anti-solar DR regime (i.e. for $\Delta \Omega < 0$), where the pole is rotating faster, mode amplitudes diminish more rapidly than in the uniform rotation case when rotation increases.
\par
\begin{figure}[h!]
    \centering
    \includegraphics[width =0.5\textwidth]{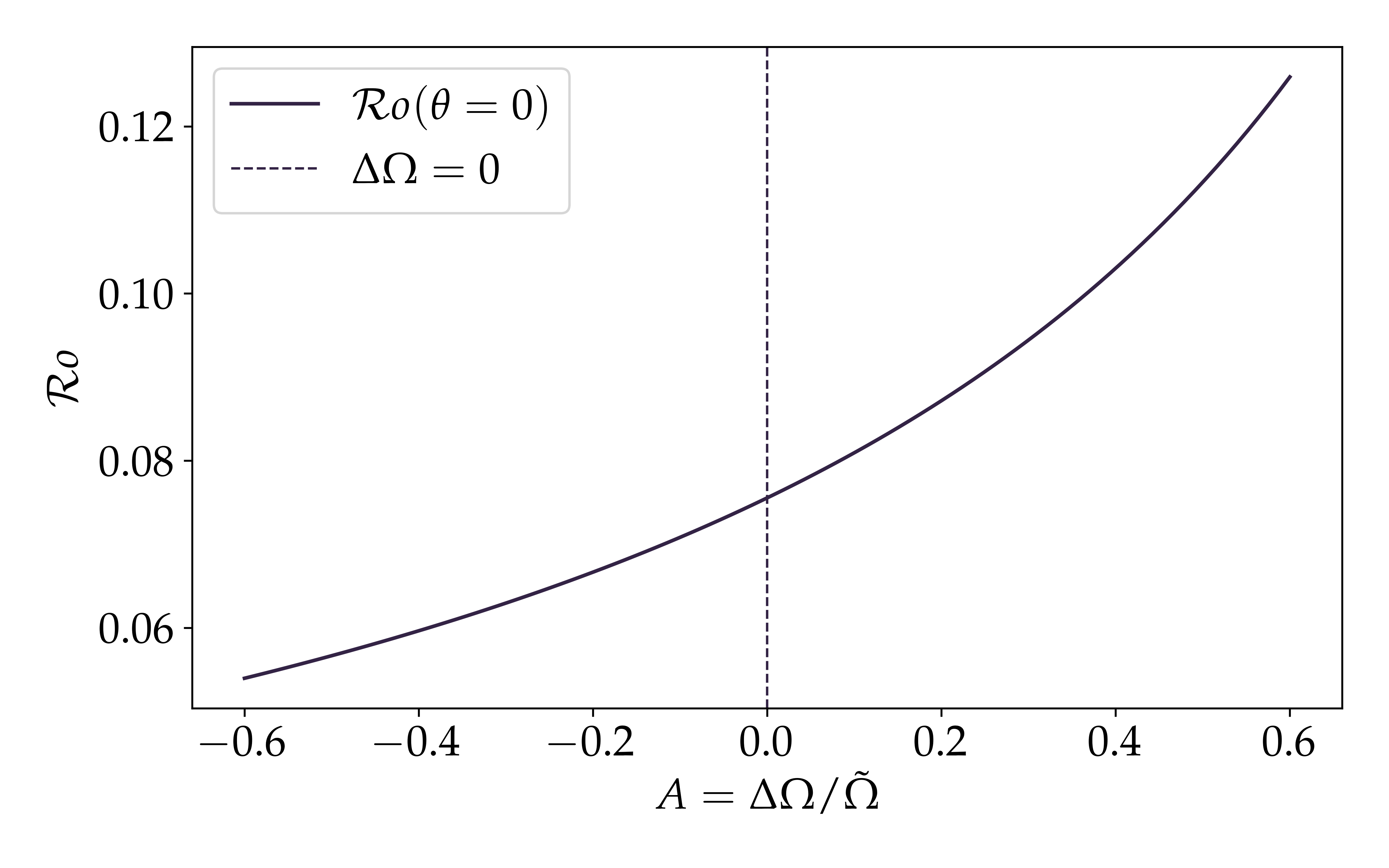}
    \caption{Rossby number computed at the pole $\theta =0$, and at radius $r = 0.8 R_{\odot}$ for $\tilde{\Omega} = \Omega_{\odot}$ as a function of the relative DR rate $A$. For negative values of $A$ (antisolar DR), the Rossby number increases more slowly compared to positive values of $A$ (solar DR profile). }
    \label{fig:rossby_diff}
\end{figure}

 Second, there is an asymmetry in this result. For the same value of $\lvert \Delta \Omega \rvert / \tilde{\Omega}$, the difference in power injected between (DR) and uniform rotation is more pronounced in solar DR regimes than in anti-solar DR regimes. This is due to the local modification of the Rossby number: for a given value of the mean rotation rate $\tilde{\Omega}$, at a fixed radius $r$ and at the pole $\theta=0$, it scales like $\mathcal{R}o \sim u_0/ \left[2\ell_0 \tilde{\Omega}(1+2A/3) \right]$. This function is not symmetric: it increases more rapidly for $A>0$ (i.e. solar DR regime) than for $A<0$ (i.e. antisolar DR regime). We illustrated this behaviour in Fig. \ref{fig:rossby_diff}. It then influences differently the convection in the R-MLT, and in turn the stochastic excitation of acoustic modes. Consequently, solar DR profiles exhibit a greater difference in the excitation power when compared to the uniform rotation case due to the more rapid variation in the Rossby number. 
\par Third, for a given $\lvert \Delta \Omega \lvert$ value, the difference between solar and antisolar differential rotation rates increases with rotation rate. For a relative DR $\lvert \Delta \Omega \rvert/\tilde{\Omega} = 1$, we find a difference of approximately 20\% between the solar and antisolar cases, for a $1 \Omega_{\odot}$ mean rotation rate.
\par Finally, Fig. \ref{fig:differential_rotation} has to be considered with caution: indeed, the relative DR is not a free parameter in our modelling, as it depends on the star's age, mass and its overall rotation rate \citep[see e.g.][]{saar_starspots_2010, brun_differential_2017, brun_powering_2022}. In particular, the anti-solar DR regime displayed in this figure is found in slowly rotating stars only, according to numerical simulations \citep[see e.g.][]{noraz_hunting_2022}.

\begin{figure}[h!]
    \centering
    \includegraphics[width =0.5\textwidth]{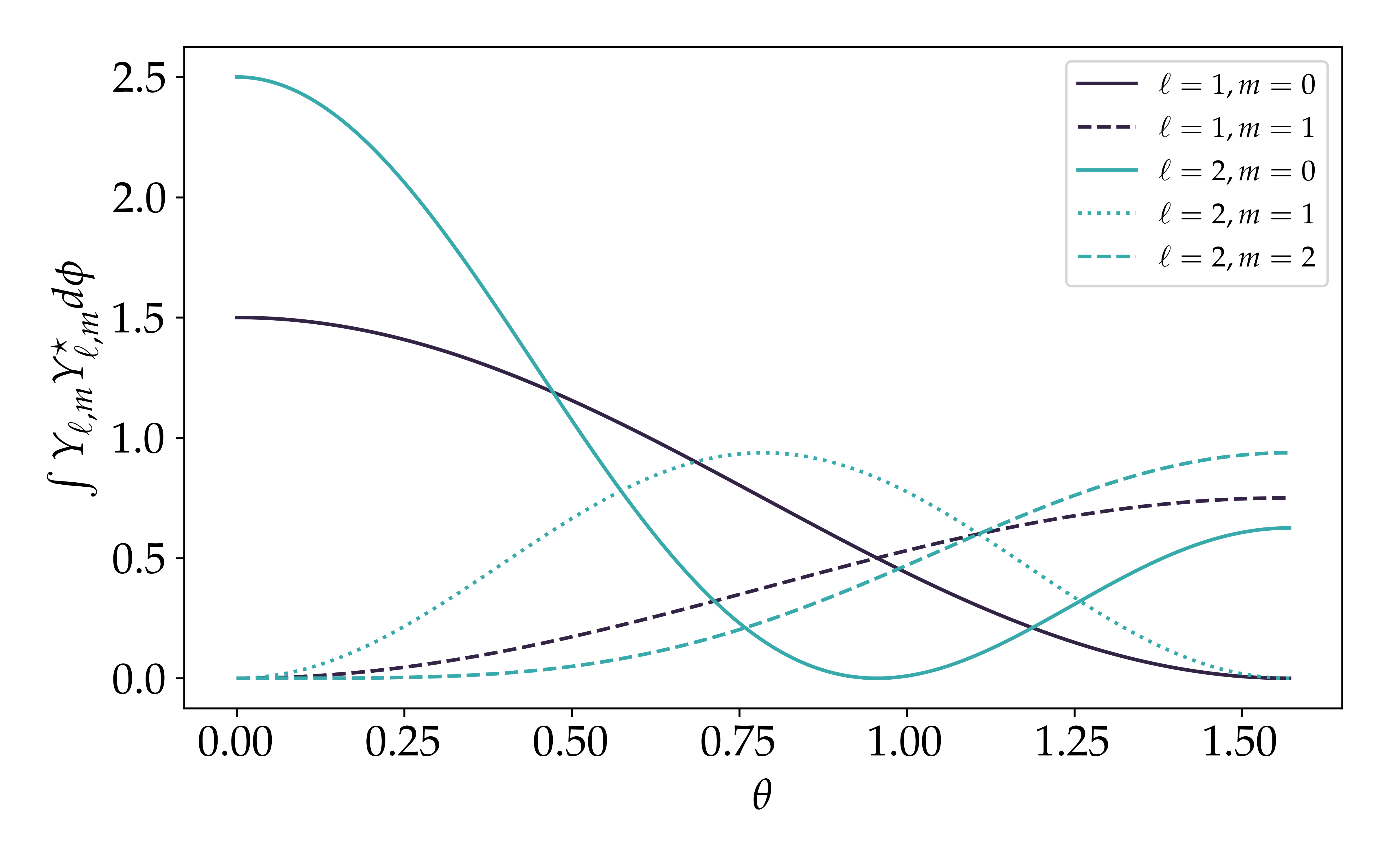}
    \caption{Azimuthal integral of the modulus of the spherical harmonics $\int_0^{2 \pi} Y_{\ell,m} Y_{\ell,m}^{\star} d\varphi$, which appears in Eq. (\ref{eq:reynolds-int}).}
    \label{fig:spherical_harmonics}
\end{figure}

As we have previously shown in \cite{bessila_impact_2024}, the power injected into the oscillations depends on the azimuthal order $m$, due to the integral over the spherical harmonics in Eq. (\ref{eq:reynolds-int}). Since the Rossby number is larger at the equator, the source term triggering the oscillations is weaker near the poles than at the equator. Axisymmetric modes, for which the spherical harmonics integral over the azimuth is maximum near the pole, thus exhibit lower amplitudes than their non-axisymmetric counterparts. We show in Fig. \ref{fig:influence_m_dr} the $\ell = 1, n = 7$ mode amplitudes as a function of the mean rotation for different values of DR and for different azimuthal orders $m$. As opposed to the trend for the $m=0$ modes, the $m=\pm 1$ mode amplitudes are more inhibited for the solar DR regime. Indeed, the integral over the spherical harmonics in Eq. (\ref{eq:reynolds-int}) gives more weight to the latitudes towards the equator than the polar region for this sectoral mode, as illustrated in Fig. \ref{fig:spherical_harmonics}. As the rotation rate at the equator is faster than the poles in the solar DR regime, the power injected into the $m=1$ mode is more inhibited. Additionally, mode amplitudes vary significantly with azimuthal order. For example, for the $\ell = 1, n=7$ mode, amplitudes with $m = \pm 1$ are about $30\%$ higher that the corresponding $m=0$ mode amplitudes, for a $10 \Omega_{\odot}$ rotation rate. This is an interesting result as it is often assumed in the literature that modes sharing the same radial order and spherical degree $\ell$ but differing in azimuthal order $m$ carry the same energy. This hypothesis of equipartition is for example used to infer stellar inclinations from solar-like oscillations \citep{gizon_determining_2003}. It would be interesting to extend such work including the effects of the azimuthal order on the mode excitation. While the present paper focuses on the injection of energy in the modes, accounting for the effects of (differential) rotation on mode damping rates will also be essential to properly model mode amplitudes. Finally, better constraining the amplitudes of acoustic modes could provide valuable constrains on DR in solar-like stars. If this effect is detectable in observations, the amplitude ratio between axisymmetric and non-axisymmetric modes could serve as a proxy to constrain differential rotation in solar-like stars. Knowing a star’s average rotation rate, measuring the relative amplitudes of modes with the same $\ell \text{ and } n)$ but varying $m$ could provide insights on the rotational shear.

\begin{figure}[h]
    \centering
        \includegraphics[width=\linewidth]{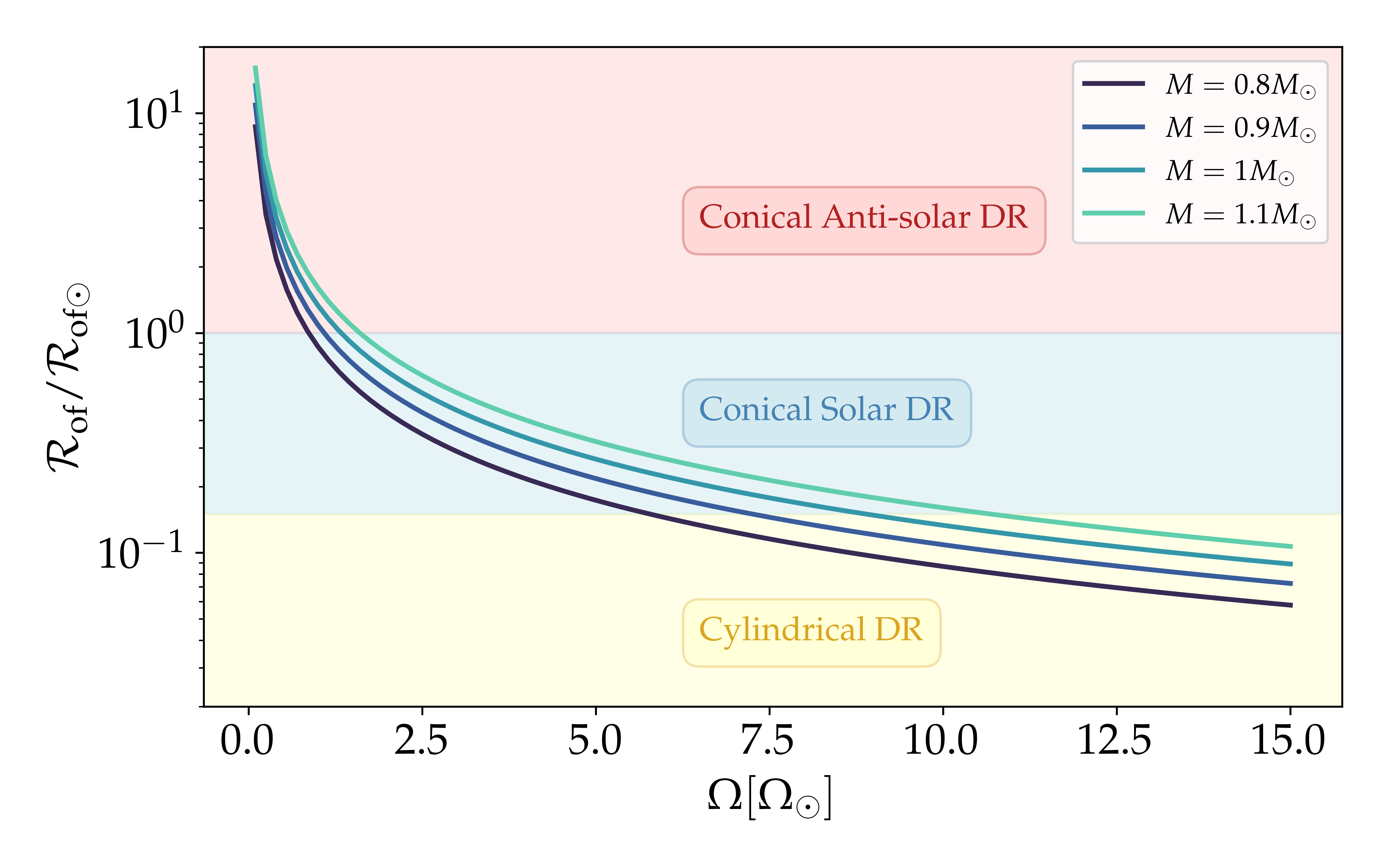}
    \includegraphics[width=\linewidth]{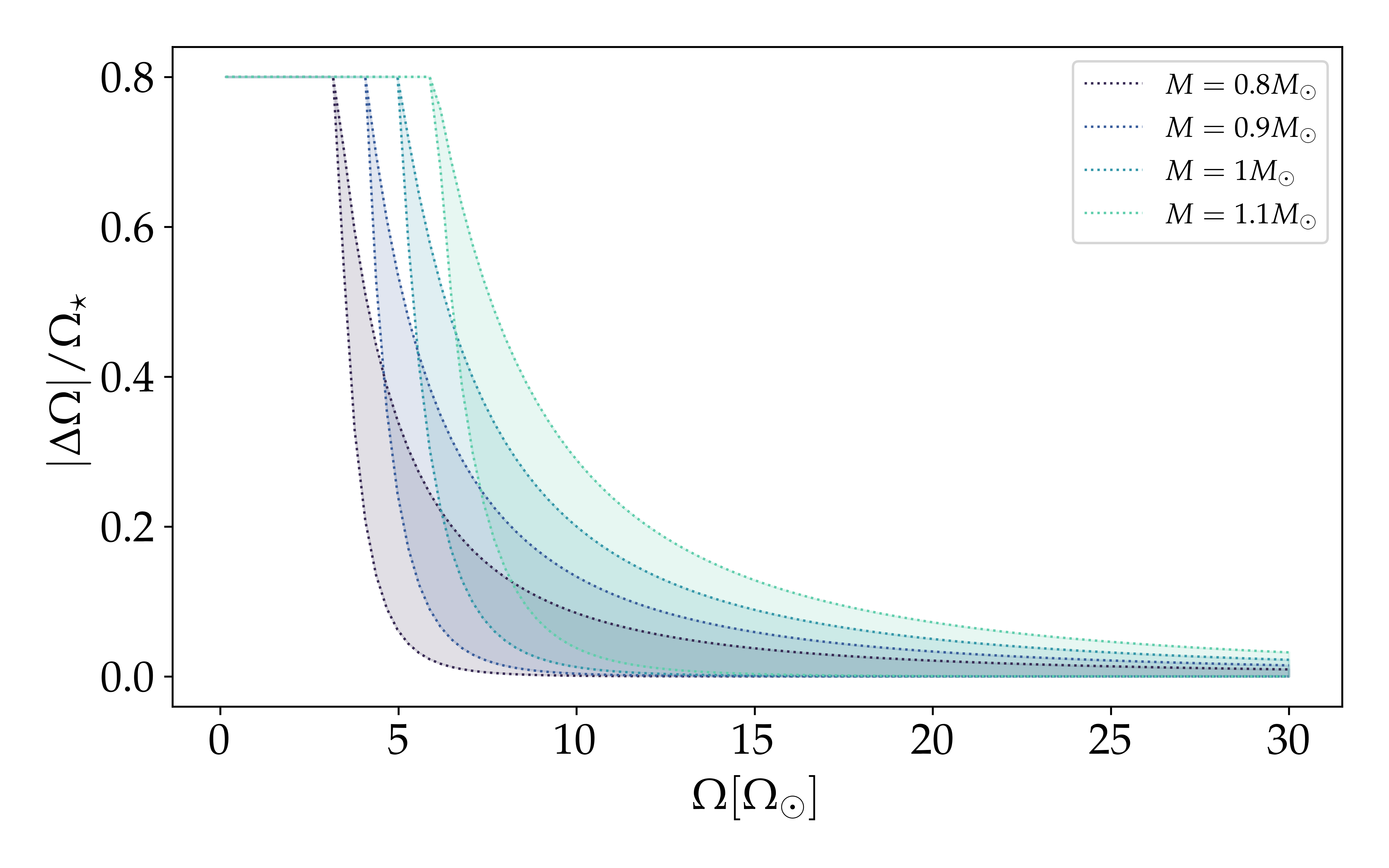}
    \caption{(Top) Fluid Rossby number according to Eq. (\ref{eq:def_rof}) from \cite{noraz_hunting_2022}. We precise here the predicted DR regime. (Bottom) Relative DR according to the scaling Eq. (\ref{eq:scaling_dr_p}) from \cite{brun_powering_2022}.  As the exponent $p$ for these scaling laws is between $p=2$ and $6$, we represent for each model the range of possible values. The choice for the exponent $p=6$ leads to a steeper diminution of the DR, when the overall rotation increases.}
    \label{fig:rossby_DR}
\end{figure}

\section{Dependance on the stellar mass and rotation}
\label{sec:dependance_mass}

\begin{figure*}[]
    \centering
    \includegraphics[width=0.7\textwidth]{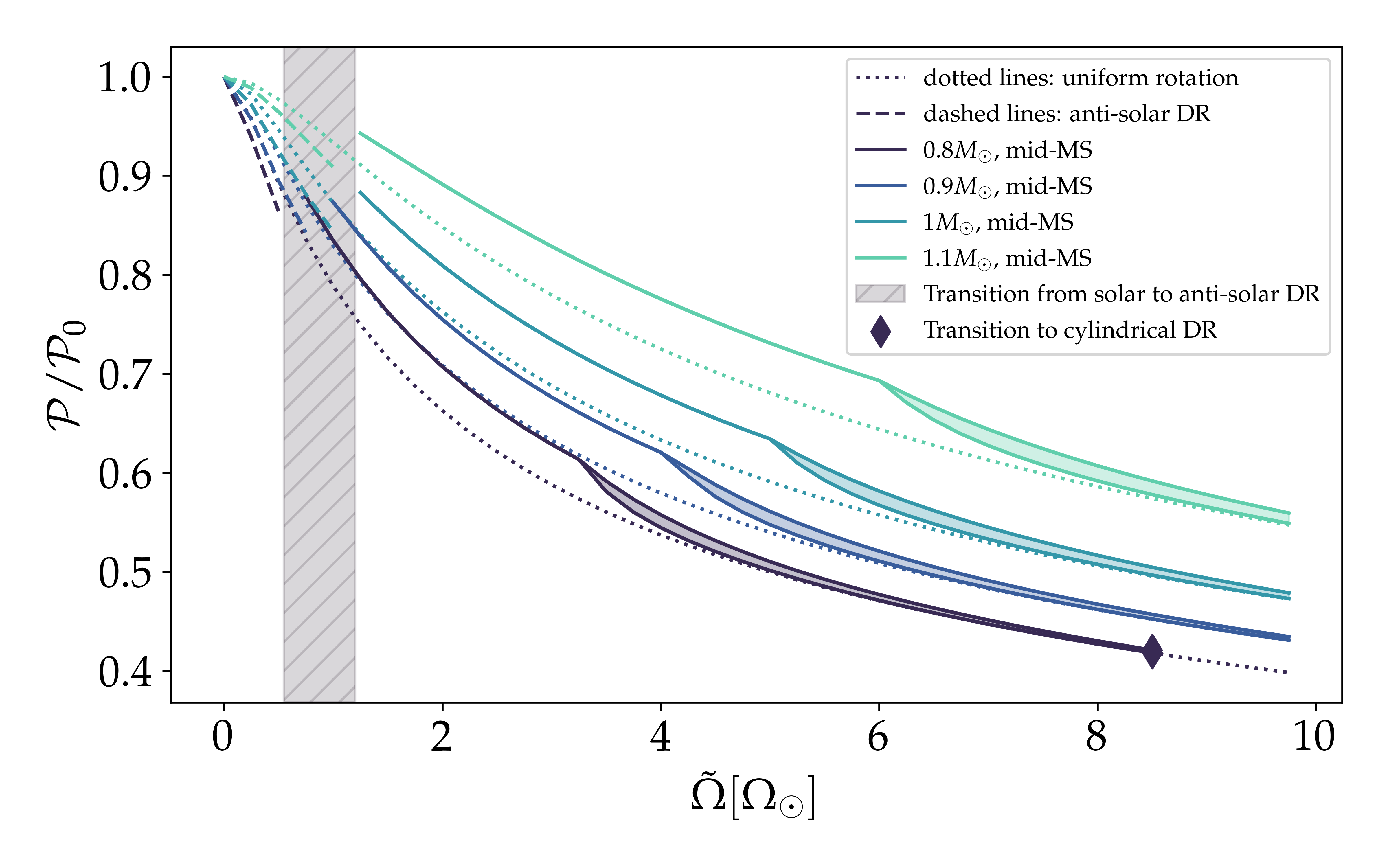}
    \caption{Relative power injected into the oscillations with rotation compared to the non-rotating case for mode ($\ell=0, n=7, m=0$), using the scalings given in Eqs. (\ref{eq:scaling_dr_p}) and (\ref{eq:def_rof}). The dashed line represents the antisolar DR regimes. As in Fig. \ref{fig:rossby_DR}, from $\tilde{\Omega} \sim 5 \Omega_{\odot}$, we propose a minimum and maximum value depending on the choice for the exponent $p$ in Eq. (\ref{eq:scaling_dr_p}), highlighted with the color zones. The dotted lines represent the uniform rotation case. The hatched grey zone highlights the transition between solar and anti-solar DR regimes. Finally, the diamond symbol marks the transition to the cylindrical DR.}
    \label{fig:dr_mass}
\end{figure*}

\label{sec:star_mass}
\par Differential rotation is not independent of the masses and rotation of the stars. Observational studies with \textit{Kepler}  underlined that the DR tends to increase when stars rotate faster \citep{reinhold_discriminating_2015}. It has been predicted that the DR scales like a power law with the rotation: $\Delta \Omega \propto \Omega_{\star}^{r}$. The observational trends vary from $r = 0.2$ to $r=0.7$, depending on the types of stars \citep{donahue_relationship_1996, reinhold_fast_2013, balona_differential_2016}. Numerical simulations also investigated the relationship between the DR and the rotation period \citep[e.g.][]{brun_differential_2017, brun_powering_2022}. The exponent $r$ was found to be $r=0.66$ for hydrodynamical simulations \citep{brun_differential_2017}, while it reduces to $r=0.46$ in the framework of magnetohydrodynamical simulations \citep{brun_powering_2022}. 
Another way to evaluate the DR is to assess the relative DR $\Delta \Omega/\Omega_{\star}$. It depends on the star's fluid Rossby number, defined as: 
\begin{equation}
    \mathcal{R}o_{\rm f} = \frac{\tilde{\omega}}{2 \Omega_{\star}},
\end{equation}
where $\tilde{\omega}$ is the root-mean-square (r.m.s.) vorticity at the middle of the convective zone. The fluid Rossby number is a direct comparison of the convective advection term and the Coriolis acceleration in the Navier–Stokes equation.
\cite{saar_starspots_2010}, followed by \cite{brun_powering_2022} noticed that the relative DR is nearly constant for $\mathcal{R}o_{\rm f} > 0.2$, and it drops for fast rotators $\mathcal{R}o_{\rm f} \leq 0.2$ following: 
\begin{equation}
\begin{aligned}
        &\lvert \Delta \Omega \rvert /\Omega_{\star} \propto \mathcal{R}o_{\rm f}^{p} \text{ if } \mathcal{R}o_{\rm f} \leq 0.2, \\
        &\lvert \Delta \Omega \rvert /\Omega_{\star} = 0.8 \text{ if } \mathcal{R}o_{\rm f} > 0.2. \\
\end{aligned}
\label{eq:scaling_dr_p}
\end{equation} 

\begin{figure}[h!]
    \centering
    \includegraphics[width=\linewidth]{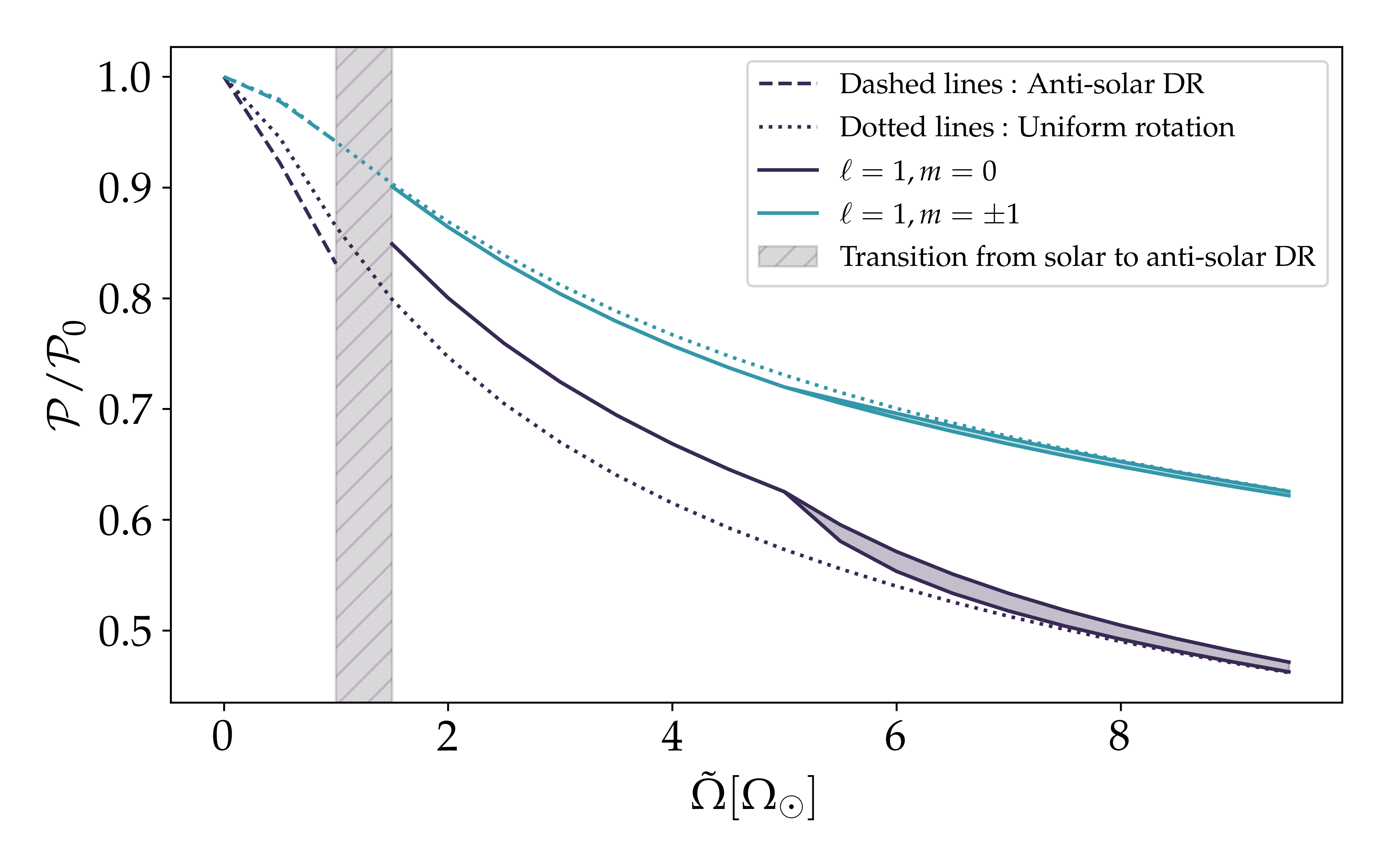}
    \includegraphics[width=\linewidth]{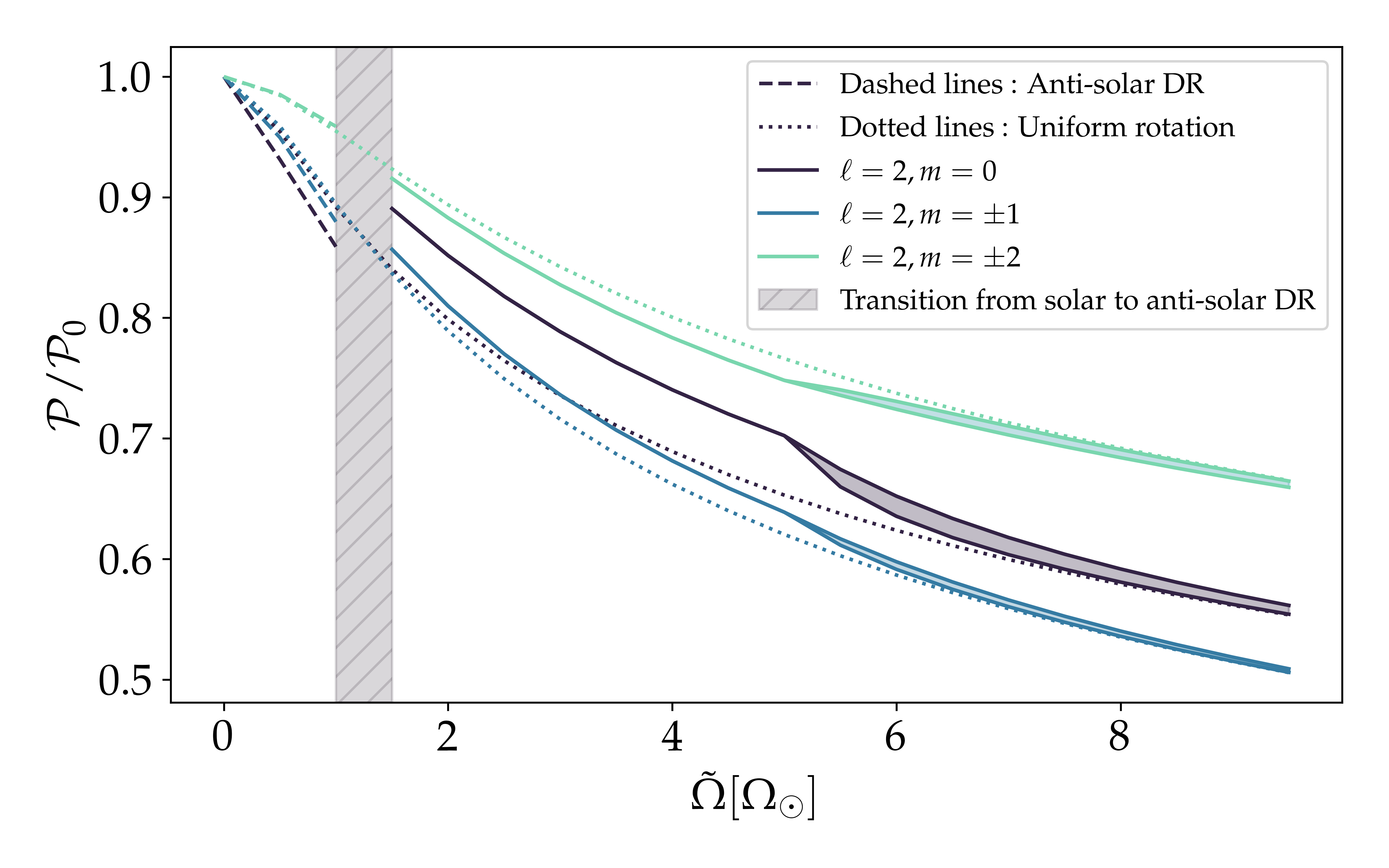}
    \caption{Influence on the angular degree and azimuthal number on the power injected by the stochastic excitation. (Top) $\ell = 1, n=7$ modes. (Bottom) $\ell = 2, n=7$ modes.}
    \label{fig:influence_m}
\end{figure}
The exponent is found to be $p=2$ in the work by \cite{saar_starspots_2010}, while \cite{brun_powering_2022} find larger values $p \in [2,6]$. 
As mentioned in the introduction, some transitions between different DR regimes were found in numerical simulations \citep{gastine_solar-like_2014, kapyla_confirmation_2014,karak_magnetically_2015, noraz_impact_2022}: for a fluid Rossby number above $\mathcal{R}o_{\rm f} \sim 1$, stars exhibit an anti-solar DR profile.

In this section, we assess the impact of DR on the stochastic excitation of modes using those prescriptions. First, we compute stellar models with masses ranging from $M = 0.8 M_{\odot}$ to $M = 1.1 M_{\odot}$ and a $Z = 0.02$ metallicity, at mid-main-sequence. Their path on the HR diagram, as well as their profiles for density, non-rotating convective velocity and convective mixing length (the inverse of the non-rotating convective wavenumber $k_0$ in our formalism) are detailed in Appendix \ref{sec:properties_stellar}. Second, we compute their fluid Rossby number depending on their rotation rate. We use the scaling proposed by \cite{noraz_hunting_2022}, in  their equation (18): 
\begin{equation}
    \mathcal{R}o_{\rm f} = \frac{\mathcal{R}o_{\rm f, \odot}}{\Omega_{\star}}\left( \frac{T_{\rm eff, 
    \star}}{T_{\rm eff \odot}}\right)^{3.29}, 
    \label{eq:def_rof}
\end{equation}
where $T_{\rm eff, \odot}$ and $\mathcal{R}o_{\rm f, \odot}$ are the effective temperature and fluid Rossby number of the Sun, respectively. The solar fluid Rossby number is between 0.6 and 0.9 according to numerical simulations \citep[e.g.][]{brun_powering_2022}. The difference arises from the convective conundrum, which reflects uncertainties in our understanding of solar convection \citep{hanasoge_solar_2015, noraz_global_2025}. In this work, we chose to take $\mathcal{R}o_{\rm f, \odot} = 0.75$ and $T_{\rm eff, \odot} = 5772~K$. We then assess the stars' relative DR rate following Eq. (\ref{eq:scaling_dr_p}), with $p \in [2,6]$ as an exponent, which gives the lower and upper limits for the relative DR rate according to \cite{brun_powering_2022}. The results for the fluid Rossby number and the relative DR according to Eqs. (\ref{eq:scaling_dr_p}) and (\ref{eq:def_rof}) are plotted in Fig. \ref{fig:rossby_DR}. In agreement with other works such as \cite{noraz_magnetochronology_2024}, the relative DR is smaller for low-mass stars at a given mean rotation rate. This is due to the fluid Rossby number being smaller for smaller stellar masses. In addition, we take the relative DR as a constant for low rotation rates i.e. for $\mathcal{R}o_{\rm f} > 0.2$, as highlighted by previous works \citep{saar_starspots_2010, brun_powering_2022}.

Next, we compute the power injected in the oscillations using the stochastic excitation formalism. For values of fluid Rossby number greater than 1, hence low rotation rates, we assume an anti-solar DR profile. We show the resulting power injected into the mode ($\ell = 0, m = 0, n = 7$), taking into account DR in Fig. \ref{fig:dr_mass}. In the anti-solar case, the power injected into the modes diminishes more rapidly when rotation increases when compared to the uniform rotation case. For a given stellar mass, when the rotation rate increases, the fluid Rossby number diminishes. When the fluid Rossby number goes below unity, around $\tilde{\Omega} \sim \Omega_{\odot}$, the rotation regime switches to a solar DR profile. From there, the power diminishes less when rotation increases when compared to the uniform rotation case, in agreement with Fig. \ref{fig:differential_rotation} where we highlighted that a solar DR makes the modes less inhibited by rotation than in the uniformly rotating case. Acoustic modes in stars with anti-solar DR would then be more difficult to detect than in the uniformly rotating case. This must be taken into account when hunting stars hosting potentially anti-solar DR \citep{noraz_hunting_2022}.

Finally, we compute the power injected in the modes with the same angular degree $\ell$, but a different azimuthal number $m$. We show the result in Fig. \ref{fig:influence_m}, for a $1 M_{\odot}$ stellar model. In this figure, all the modes are not equally influenced by rotation. For the sectoral modes with $m = \pm \ell$, taking into account solar DR results in more inhibited mode amplitudes in the solar DR regime, as opposed to what we witnessed for the other modes. As explained in Sec. \ref{sec:dependance_mass}, for the sectoral modes the azimuthal integral of the spherical harmonics modulus in Eq. (\ref{eq:reynolds-int}) gives more weight to equatorial amplitudes, resulting in less power injected into the acoustic modes than for the uniform rotation case (see Fig. \ref{fig:spherical_harmonics} and \ref{fig:influence_m}). Moreover, we find the same trend for less massive solar-like stars (with $M = 0.8 M_{\odot}$ and for more massive solar-like stars (with $M = 1.2 M_{\odot}$), as shown in Appendix \ref{sec:appendix_m} in Fig. \ref{fig:influence_dr_mass_l_1} and \ref{fig:influence_dr_mass_l_2}.

\section{Conclusion and perspectives}
\label{sec:conclusion}

In this work, we study for the first time the impact of DR on acoustic mode excitation by turbulent convection in pulsating main-sequence solar-like stars, taking into account a conical DR profile in their convective zone. We extend the theoretical formalism derived by \cite{samadi_excitation_2001, belkacem_mode_2009} and \cite{bessila_impact_2024} and we consider that turbulent convection is locally modified by DR. We model this modification using the Rotating Mixing-Length theory \citep{stevenson_turbulent_1979, augustson_model_2019}, a single-mode approach of convection, which predicts that rapid rotation modifies convection: the convective velocity and the characteristic turbulent eddy size are then diminished when rotation increases.

In this framework, we find in general that the higher the rotation rate, the lower the mode amplitudes. More specifically, an anti-solar DR leads to a stronger diminution compared to the uniformly rotating case. Conversely, a solar DR profile leads to a slower diminution when the overall rotation rate increases compared to the uniform rotation case. We find an asymmetry in this tendency: for the same value of $\lvert \Delta \Omega \rvert$, modes are more affected by rotation for a solar DR. Next, we apply these results to compute the power injected by the stochastic excitation for mid-main sequence solar-like stars with masses ranging from $0.8$ to $1.1 M_{\odot}$. We used the scaling provided by \cite{saar_starspots_2010, brun_powering_2022, noraz_hunting_2022} to model the variation of the DR with stellar mass and rotation based on state-of-the-art observations and large-eddy numerical simulations of the dynamics of differentially rotating convective envelope of low-mass stars in spherical geometry, both in the hydrodynamical and magnetohydrodynamical cases. Such relations are suitable for qualitative estimates, although we must keep in mind that reality is more complex. We find that the power injected into the modes diminishes more slowly for values of $\Omega \approx 1 \Omega_{\odot}$ when the stars are predicted to transition from an anti-solar to a solar DR profile. 
We have also studied the influence of the azimuthal order $m$, wich leads to a variation in the mode amplitudes, as highlighted before for the uniform rotation case \citep{bessila_impact_2024}. It would be interesting to take this result into account when inferring stellar inclinations for observations and go beyond the equipartition hypothesis \citep{gizon_determining_2003}. We have also shown that for the sectoral modes, with $\ell = \pm m$, the solar DR regime results in more inhibited amplitudes that for the uniform rotation case.
These results highlight that the anti-solar rotation regime observed in numerical simulations might be more difficult to detect with asteroseismology, as acoustic modes are less excited when compared to the uniform rotation case at a fixed mean rotation. This outcome will be useful to predict the mode detection probability in the preparation of the \textit{PLATO} mission \citep{rauer_plato_2014, rauer_plato_2024, goupil_predicted_2024}. However, mode amplitudes are a balance between driving and damping \citep[see e.g.][]{samadi_stellar_2015}. To correctly assess the detection probability, one must also model the damping with differential rotation. This is not investigated in the present paper and is left for future work. Additionally, evaluating the impact of metallicity on the relative variation of differential rotation \citep[see e.g.][]{see_photometric_2021, noraz_hunting_2022}, and consequently on stochastic excitation, would provide further insights.

To go further, deriving a model with both magnetic field and DR is crucial. Magnetic fields are ubiquitous in solar-like stars and are known to reduce the amplitudes of stochastically excited acoustic modes \citep{bessila_stochastic_2024}, in agreement with observational trends reported by \cite{garcia_corot_2010, chaplin_predicting_2011}, and \cite{mathur_revisiting_2019}. In late-type stars, magnetic fields are generated through dynamo action within their convective envelopes. This process involves a complex interplay between latitudinal shear and magnetic field such as the $\Omega$-effect which is one of the key mechanisms driving this dynamo \citep{charbonneau_solar_2014, brun_powering_2022}. Consequently, DR plays a critical role in understanding the generation of magnetic fields and the redistribution of angular momentum as a back reaction in the convective envelope of low-mass stars. For these reasons, our model of the stochastic excitation of acoustic modes must be extended to tackle the combined impact of DR and magnetic fields. This would allow more accurate predictions of acoustic mode amplitudes within a more general and realistic framework.

\begin{acknowledgements}
L.B. and S.M. thank the referee for their comments that considerably improved the manuscript. L.B. and S.M. acknowledge support from the  European  Research Council  (ERC)  under the  Horizon  Europe programme  (Synergy  Grant agreement 101071505: 4D-STAR), from the CNES SOHO-GOLF and PLATO grants at CEA-DAp, and from PNPS (CNRS/INSU). While partially funded by the European Union, views and opinions expressed are however those of the authors only and do not necessarily reflect those of the European Union or the European Research Council. Neither the European Union nor the granting authority can be held responsible for them.
\end{acknowledgements}

\bibliographystyle{aa}
\bibliography{references-4}

\appendix

\section{Properties of the computed stellar models}
\label{sec:properties_stellar}
\begin{figure}[h!]
    \centering
    \includegraphics[width=\linewidth]{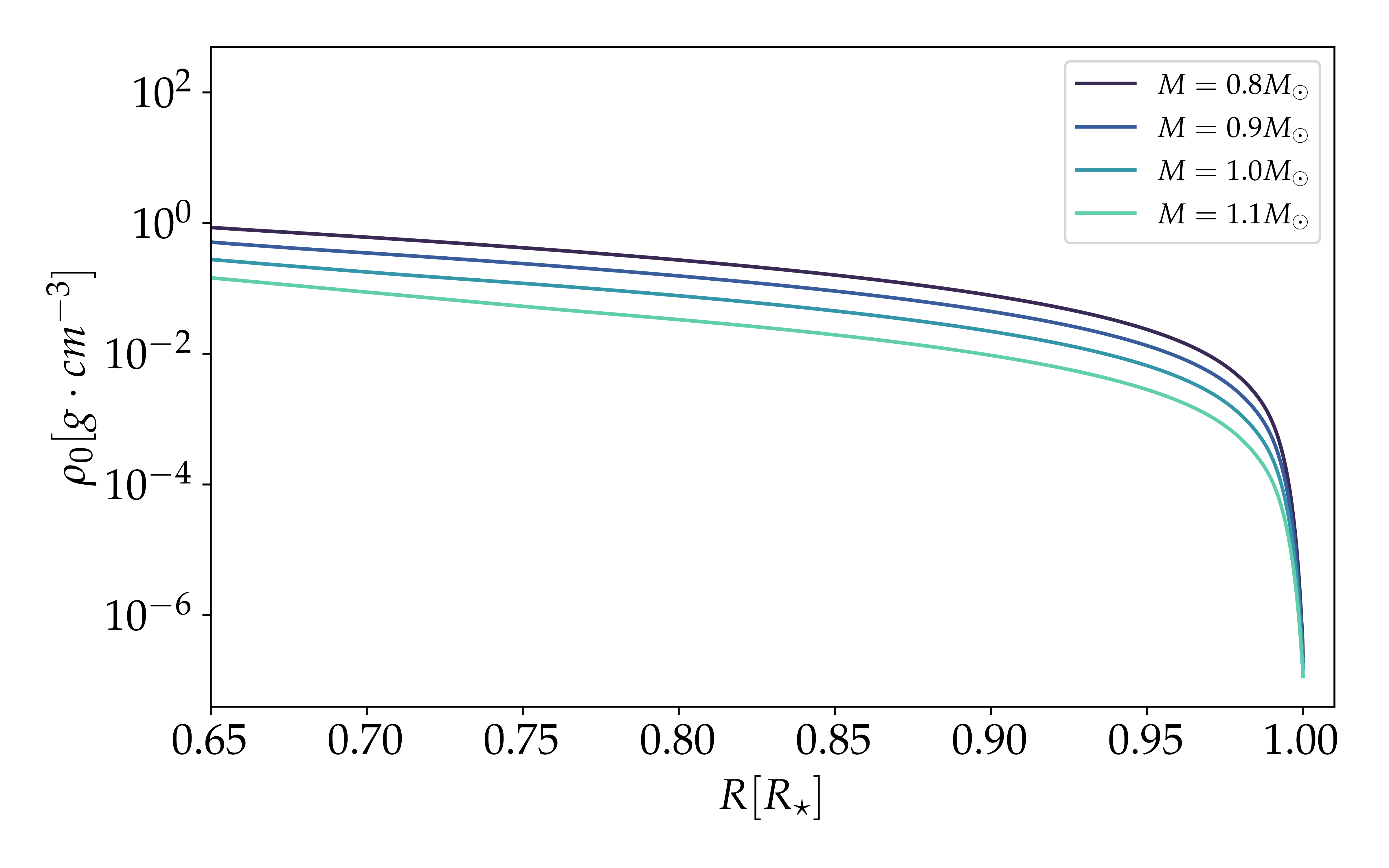}
    \includegraphics[width=\linewidth]{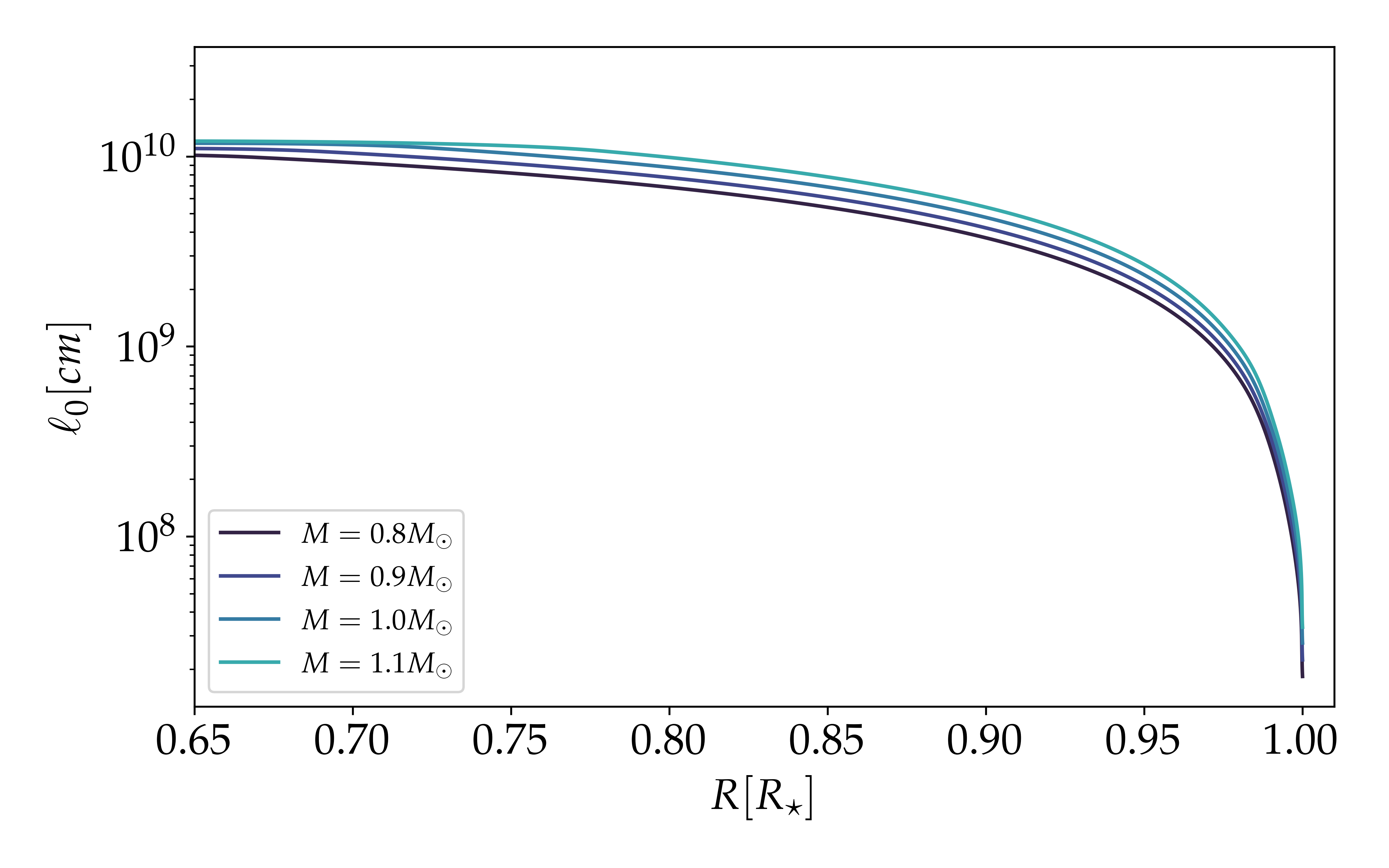}
    \includegraphics[width=\linewidth]{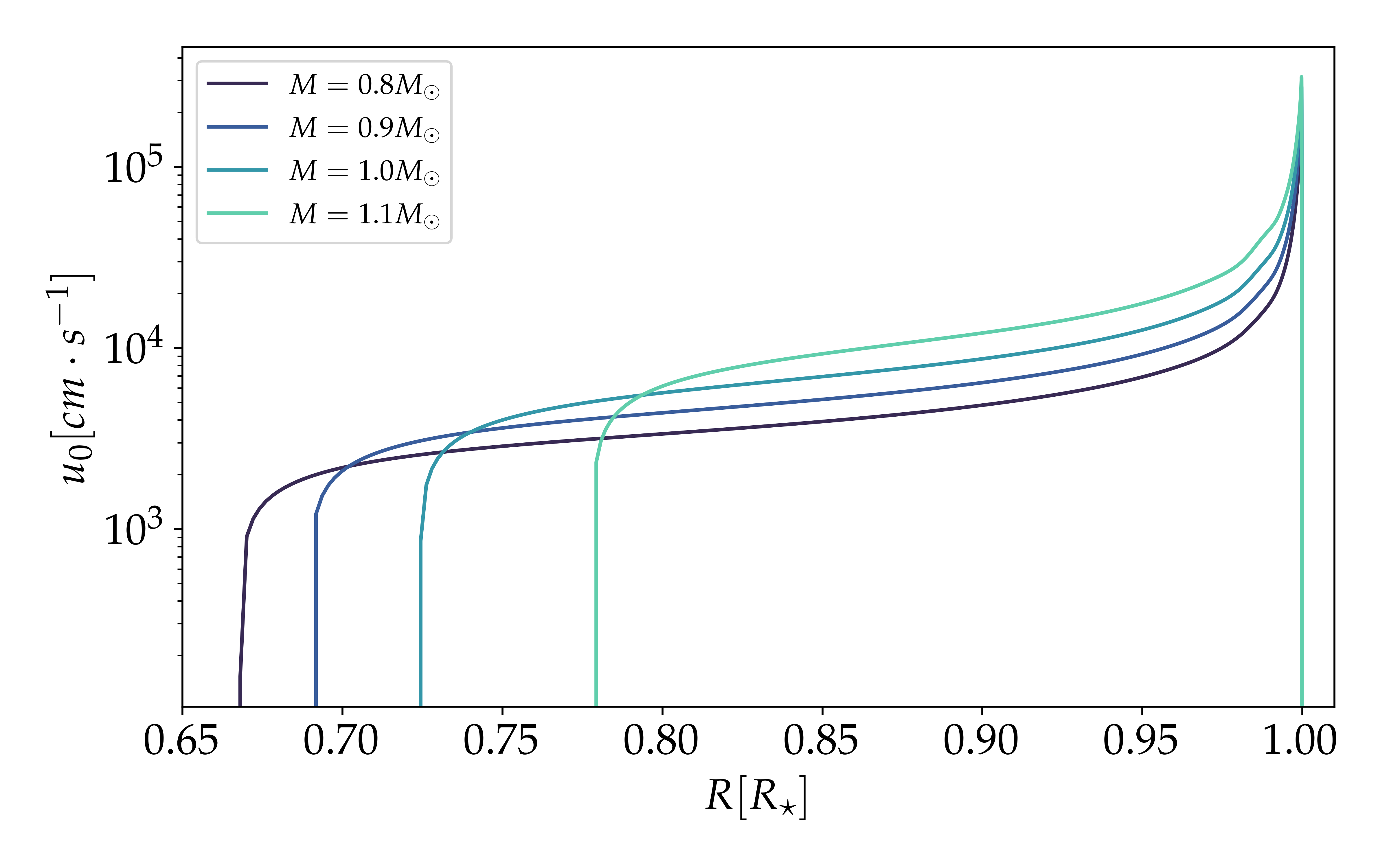}
    \caption{Density profile $\rho_0$  (top) and non-rotating convective characteristic mixing-length $\ell_0$ (middle) and velocity $u_0$ (bottom). Note that the radius $r$ is normalised to each stellar radius $R_{\star}$.}
    \label{fig:models_profiles}
\end{figure}

\begin{figure}[h!]
    \centering
    \includegraphics[width=\linewidth]{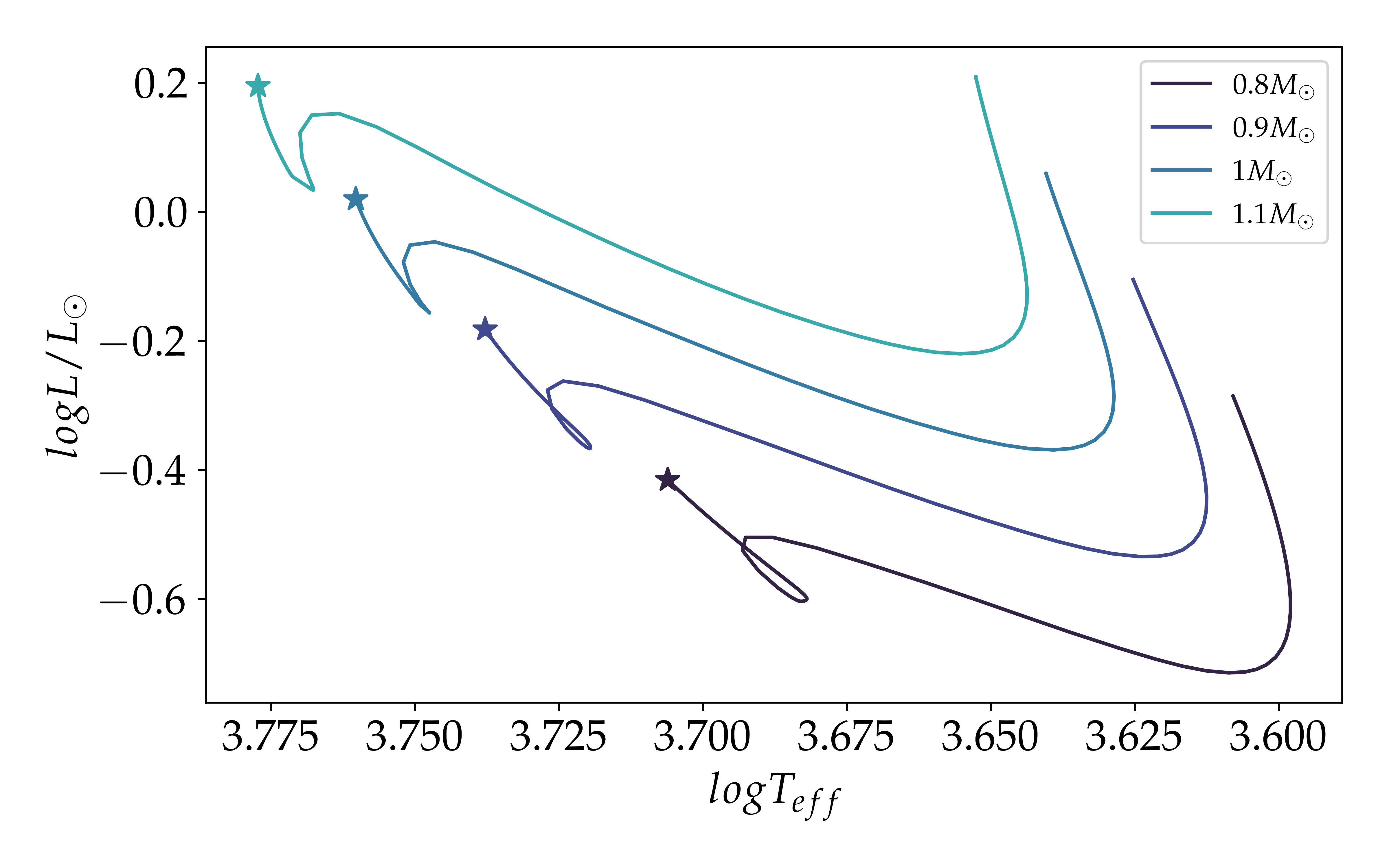}
    \caption{Position of the different stellar models in the Hertzsprung-Russell diagram.}
    \label{fig:models_hr}
\end{figure}

We detail in this section the properties of the stellar models used in Sec. \ref{sec:star_mass}. The radial profiles of the density and the non-rotating convective characteristic mixing-length and velocity are displayed in Fig. \ref{fig:models_profiles}. The corresponding HR diagrams are plotted in Fig. \ref{fig:models_hr}. We study models with different stellar masses from $0.8 M_{\odot}$ to $1.1 M_{\odot}$, at a metallicity $Z=0.02$, at mid-Main Sequence. 

\section{Influence of the azimuthal number for different stellar models}
\label{sec:appendix_m}
In this section, we compute the power injected by the stochastic excitation for stellar models with $M=0.8 M_{\odot}$ and $M = 1.1 M_{\odot}$. We retrieve the same trend than in Fig. \ref{fig:dr_mass}: in the models with a higher mass, the modes are less inhibited by the rotation. 

\begin{figure}
    \centering
    \includegraphics[width=\linewidth]{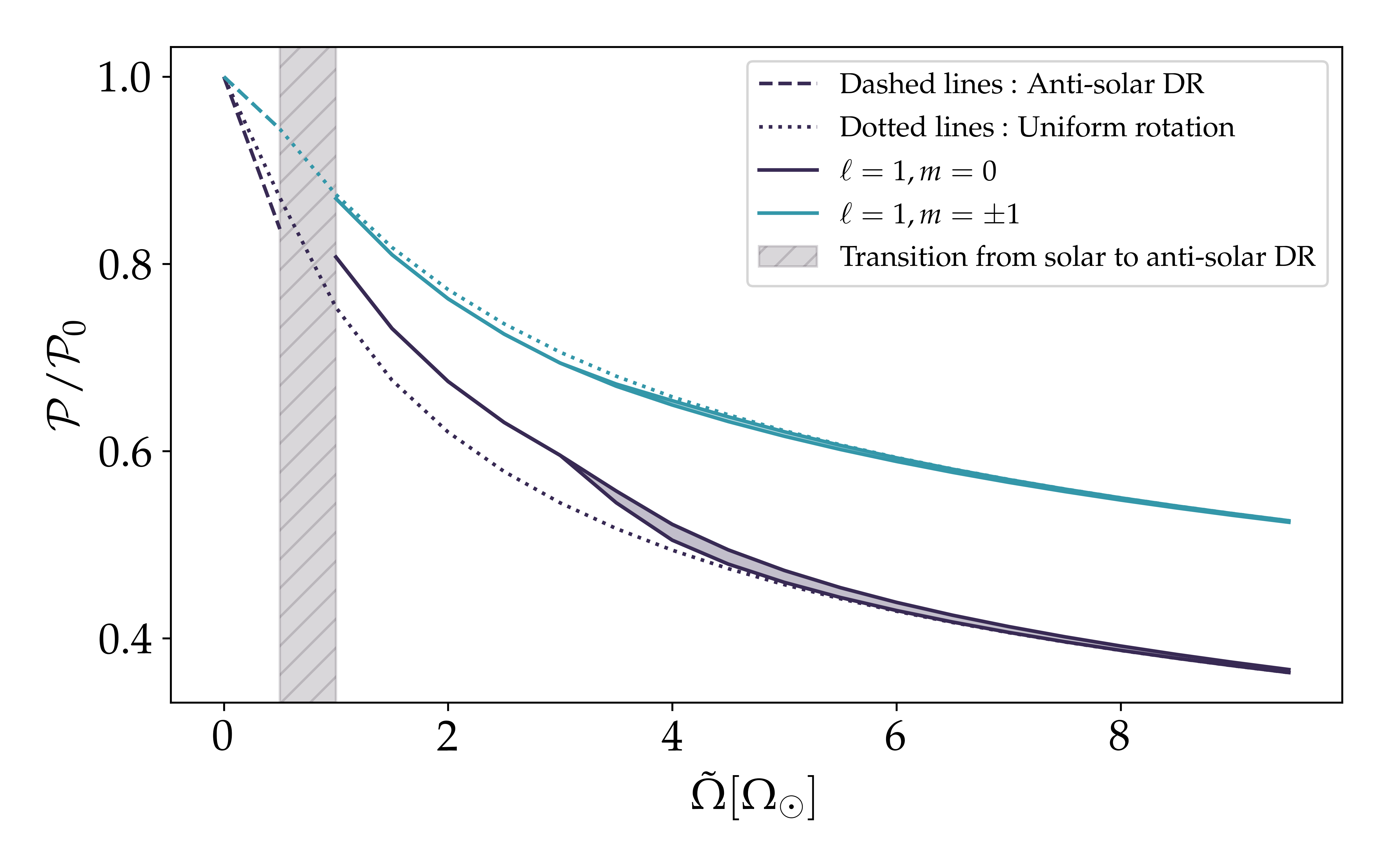}
    \includegraphics[width=\linewidth]{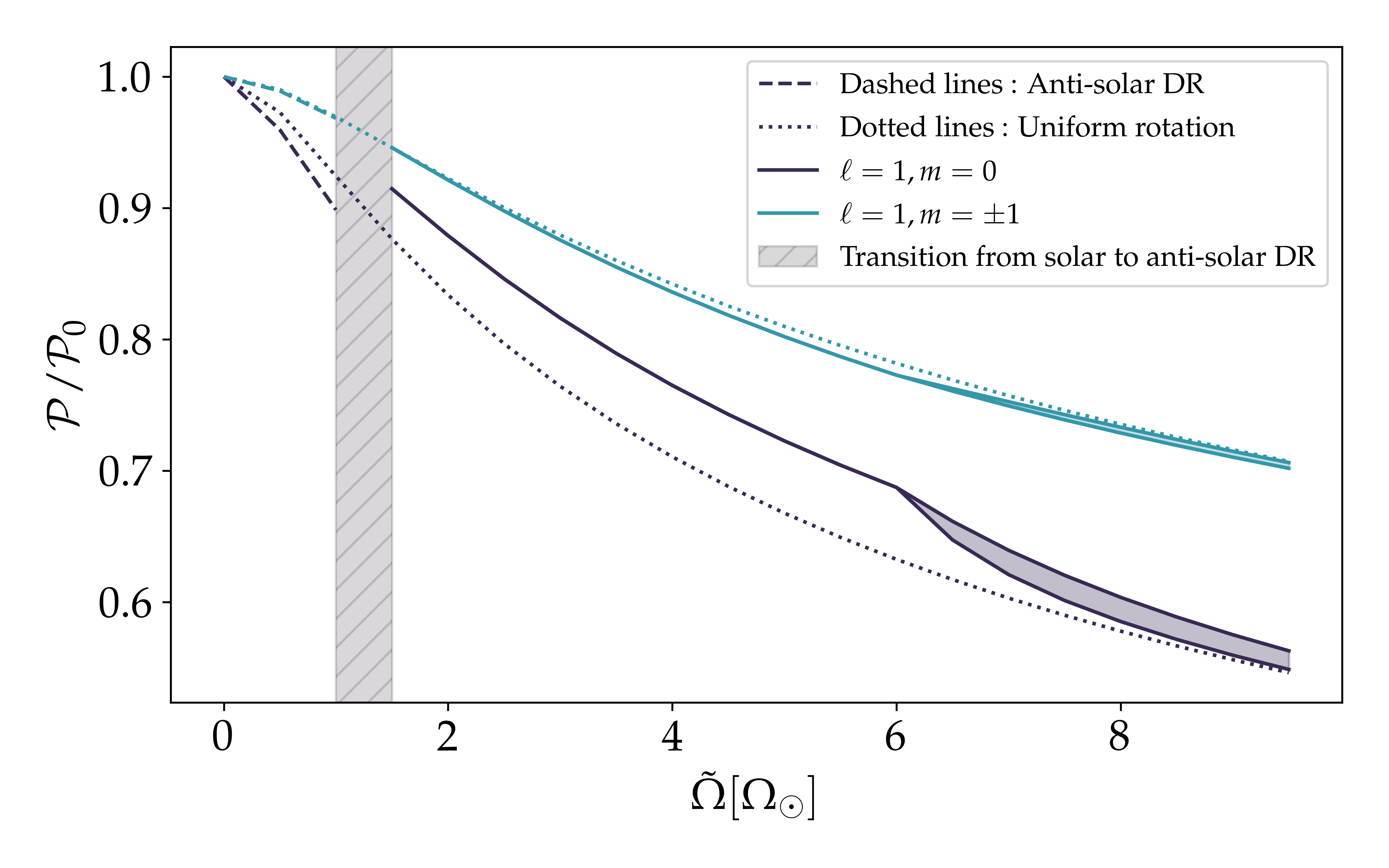}
    \caption{Influence of the differential rotation on the stochastic excitation of the modes $\ell = 1$, $n=7$ and different values of $m$. (Top) $0.8 M_{\odot}$ model (Bottom) $1.1 M_{\odot}$ model.}
    \label{fig:influence_dr_mass_l_1}
\end{figure}

\begin{figure}
    \centering
    \includegraphics[width=\linewidth]{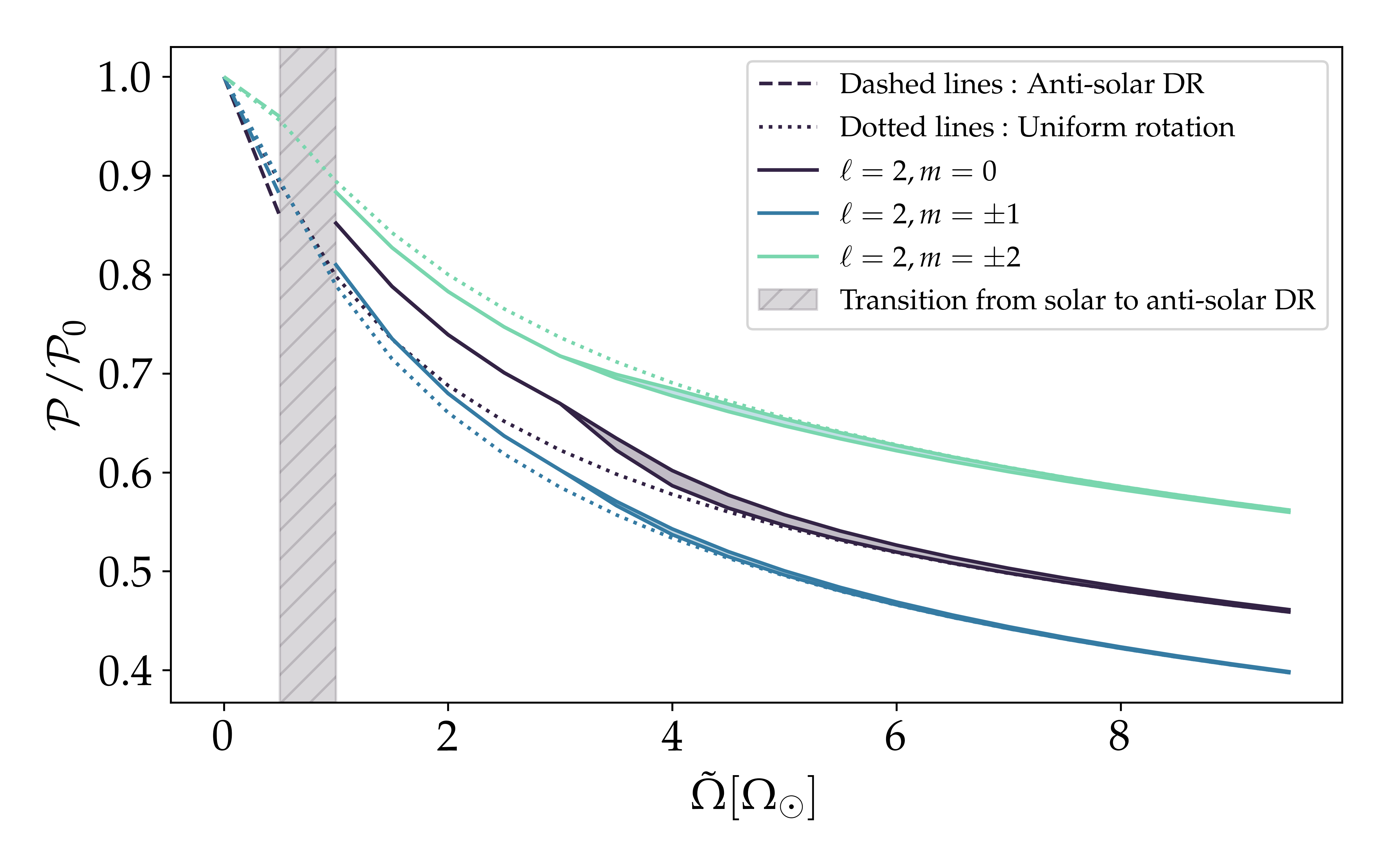}
    \includegraphics[width=\linewidth]{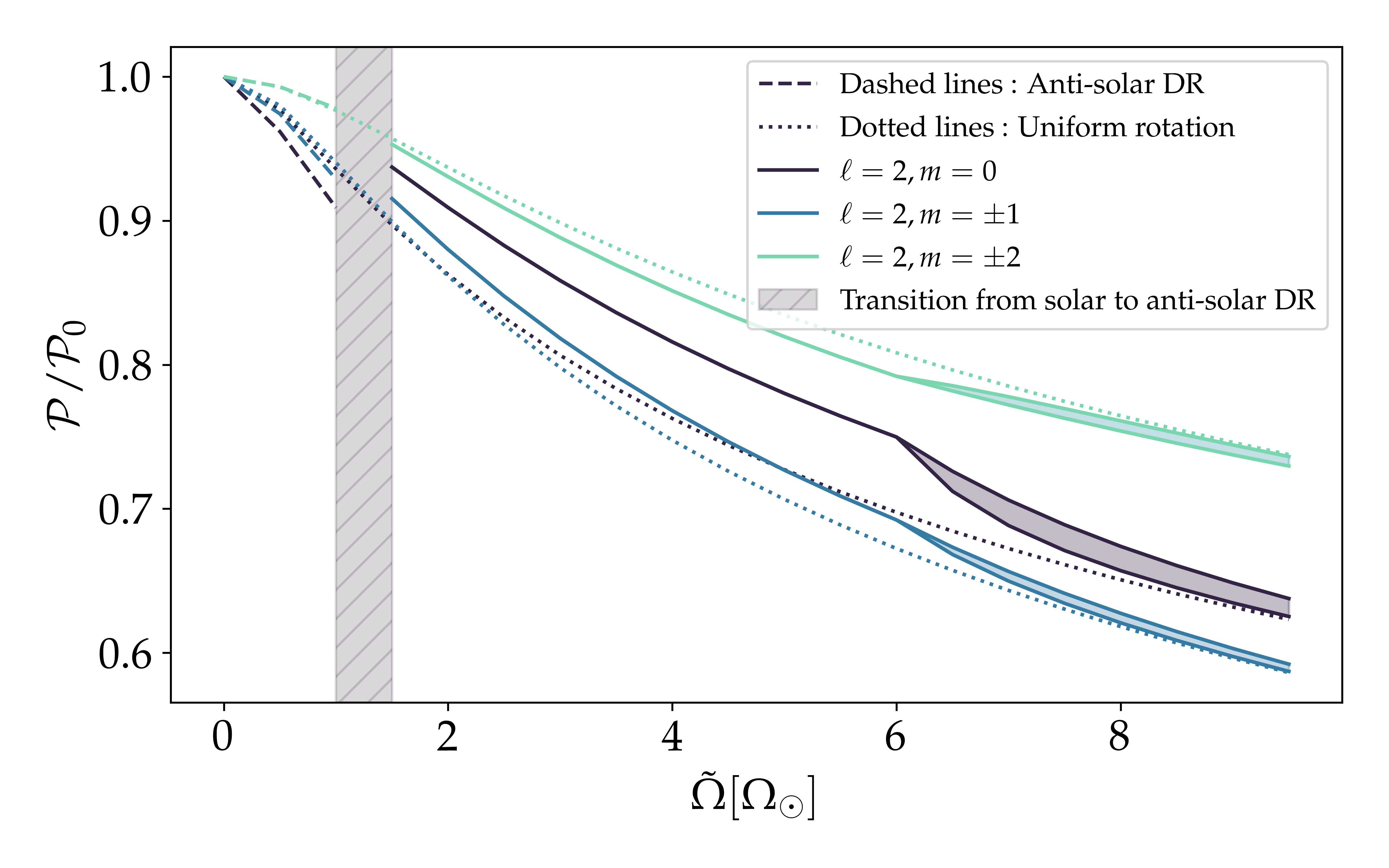}
    \caption{Influence of the differential rotation on the stochastic excitation of the modes $\ell = 2$, $n=7$ and different values of $m$. (Top) $0.8 M_{\odot}$ model (Bottom) $1.1 M_{\odot}$ model.}
    \label{fig:influence_dr_mass_l_2}
\end{figure}

\newpage
\section{MESA Inlists for the solar model}
\label{sec:inlist}
\begin{verbatim}
&kap
  ! kap options
  ! see kap/defaults/kap.defaults
  use_Type2_opacities = .true.
    Zbase = 0.016

/ ! end of kap namelist

&controls
      initial_mass = 1.0 

      ! MAIN PARAMS
      mixing_length_alpha = 1.9446893445
      initial_z = 0.02 ! 0.04, 0.002, 0.026
      do_conv_premix = .true.
      use_Ledoux_criterion = .true.

      ! OUTPUT
      max_num_profile_models = 100000
      profile_interval = 300
      history_interval = 1
      photo_interval = 300

      ! WHEN TO STOP
      xa_central_lower_limit_species(1) = 'h1'
      xa_central_lower_limit(1) = 0.01
      max_age = 6.408d9

      ! RESOLUTION
      mesh_delta_coeff = 0.5
      time_delta_coeff = 1.0

      ! GOLD TOLERANCES
      use_gold_tolerances = .true.
      use_gold2_tolerances = .true.
      delta_lg_XH_cntr_limit = 0.01
      min_timestep_limit = 1d-1

      !limit on magnitude 
      delta_lgTeff_limit = 0.25 ! 0.005
      delta_lgTeff_hard_limit = 0.25 ! 0.005
      delta_lgL_limit = 0.25 ! 0.005

      ! asteroseismology
      write_pulse_data_with_profile = .true.
      pulse_data_format = 'FGONG'
      ! add_atmosphere_to_pulse_data = .true.


/ ! end of controls namelist

\end{verbatim}

\end{document}